\documentclass[11pt]{article}
\usepackage{epsf,amsmath,amssymb,graphicx,scalefnt,rotating,enumitem,fancyhdr}
\usepackage{cite}
\usepackage{verbatim}
\usepackage{mathtools}
\usepackage[english]{babel}
\usepackage{longtable}
\usepackage[final]{showkeys}
\usepackage{slashed}
\usepackage{acronym}

\usepackage{filemod}
\usepackage[pdftex, colorlinks=true, linkcolor=black, filecolor=black,
  citecolor=black, urlcolor=black, draft=false, bookmarks,
  bookmarksnumbered=true, plainpages=false,
  linktocpage={true}]{hyperref}
\usepackage[affil-it]{authblk}
\usepackage[usenames,dvipsnames]{color}
\usepackage{tensor}
\usepackage{braket}
\usepackage[centertableaux, smalltableaux]{ytableau}
\usepackage{booktabs}
\usepackage{listings}
\usepackage[capitalize]{cleveref}

\crefname{relation}{relation}{relations}
\creflabelformat{relation}{#2{\upshape(#1)}#3} 
\crefname{type}{type}{types}
\creflabelformat{type}{#2{\upshape(#1)}#3} 

\newcounter{notecount}

\newcommand{\citere}[1]{Ref.\cite{#1}}
\newcommand{\citeres}[1]{Refs.\cite{#1}}

\newcommand{\abbrev}[1]{{\scalefont{.9}#1}}

\newcommand{\dd}{\mathrm{d}}
\newcommand{\deriv}[3]{\frac{\partial\ifthenelse{\equal{#1}{}}{}{^{#1}}%
    #2}{\partial #3\ifthenelse{\equal{#1}{}}{}{^{#1}}}}
\newcommand{\dderiv}[3]{\frac{\dd\ifthenelse{\equal{#1}{}}{}{^{#1}}%
    #2}{\dd #3\ifthenelse{\equal{#1}{}}{}{^{#1}}}}

\newcommand{\ytabdots}{\none[\raisebox{.1em}{$\scriptscriptstyle{\dots}$}]}
\newcommand{\ytabvdots}{\none[\mathbin{\raisebox{.15em}{\rotatebox[origin=c]{90}{$\scriptscriptstyle{\dots}$}}}]}

\DeclareMathOperator{\U}{U}
\DeclareMathOperator{\SU}{SU}
\DeclareMathOperator{\SL}{SL}

\newcommand{\myacrodef}[3]{\acrodef{#2}{#3}\newcommand{#1}{\ac{#2}}}
\myacrodef{\gpl}{GPL}{\abbrev{GNU} General Public License}
\myacrodef{\eft}{EFT}{Effective Field Theory}
\acrodefplural{EFT}{Effective Field Theories}
\newcommand{\efts}{\acp{EFT}}
\myacrodef{\smeft}{SMEFT}{Standard Model Effective Field Theory}
\myacrodef{\grsmeft}{GRSMEFT}{General Relativity$\oplus$\smeft}
\myacrodef{\ibp}{IbP}{integration-by-parts}
\myacrodef{\eom}{EoM}{equation-of-motion}
\acrodefplural{EoM}{equations-of-motion}
\newcommand{\eoms}{\acp{EoM}}
\myacrodef{\ssyt}{SSYT}{Semi-Standard Young Tableau}
\acrodefplural{SSYT}[SSYTx]{Semi-Standard Young Tableaux}
\newcommand{\ssytx}{\acp{SSYT}}
\myacrodef{\syt}{SYT}{Standard Young Tableau}
\acrodefplural{SYT}[SYTx]{Standard Young Tableaux}

\myacrodef{\ir}{IR}{infrared}
\myacrodef{\uv}{UV}{ultraviolet}
\myacrodef{\lwrs}{LR}{Littlewood-Richardson}
\myacrodef{\qcd}{QCD}{Quantum Chromo Dynamics}
\myacrodef{\lhc}{LHC}{Large Hadron Collider}
\myacrodef{\lo}{LO}{leading order}
\myacrodef{\nlo}{NLO}{next-to-leading order}
\myacrodef{\nnlo}{NNLO}{next-to-next-to-leading order}
\myacrodef{\llog}{LL}{leading logarithmic}
\myacrodef{\nll}{NLL}{next-to-leading logarithmic}
\myacrodef{\nnll}{NNLL}{next-to-next-to-leading logarithmic}
\myacrodef{\pdf}{PDF}{parton density function}
\myacrodef{\sm}{SM}{Standard Model}
\myacrodef{\bsm}{BSM}{beyond-the-\ac{SM}}
\myacrodef{\mssm}{MSSM}{Minimal Supersymmetric \ac{SM}}
\myacrodef{\susy}{SUSY}{Supersymmetry}
\myacrodef{\dreg}{DREG}{Dimensional Regularization}
\myacrodef{\dred}{DRED}{Dimensional Reduction}
\myacrodef{\emt}{EMT}{energy-momentum tensor}

\definecolor{codegreen}{rgb}{.58,.69,.5}
\definecolor{codegray}{rgb}{.5,.5,.5}
\definecolor{codeblue}{rgb}{.35,.39,.6}
\definecolor{codered}{rgb}{.55,.3,.45}
\definecolor{backcolour}{rgb}{.95,.95,.95}
\newcommand\YAMLkeystyle{\color{codeblue}}
\newcommand\YAMLvaluestyle{\color{black}\ttfamily\scriptsize}
\lstdefinestyle{yaml}{
    backgroundcolor=\color{backcolour},
    commentstyle=\color{codegreen},
    keywordstyle=\YAMLkeystyle,
    numberstyle=\color{codegray}\ttfamily\tiny,
    stringstyle=\color{codeblue},
    basicstyle=\YAMLvaluestyle,
    breakatwhitespace=false,
    breaklines=true,
    postbreak=\mbox{\textcolor{codegray}{$\hookrightarrow$}\space},
    captionpos=b,
    keepspaces=true,
    numbers=left,
    numbersep=5pt,
    showspaces=false,
    showstringspaces=false,
    showtabs=false,
    tabsize=2,
    comment=[l]{\#},
    keywords={version,type,generations,n_terms,n_operators,invariants,permutation_symmetries,vector,symmetry,matrix},
    moredelim=**[il][\color{codered}{:}\YAMLvaluestyle]{:},
}

\newcommand{\RHheaderline}{\textsf{TTK-22-47 / P3H-23-031 --- May 2023}}
\fancypagestyle{firstpage}
{
  
  \fancyhead[R]{\RHheaderline}
}
\allowdisplaybreaks
\title{Standard Model Effective Field Theory\\ up to Mass
  Dimension~12}
\author{R.V.~Harlander, T.~Kempkens, and M.C.~Schaaf}
\affil{Institute for Theoretical Particle Physics and Cosmology,\\ RWTH Aachen
  University, 52056 Aachen, Germany} \date{}
\begin{document}
\maketitle
\thispagestyle{firstpage}

\begin{abstract}
  We present a complete and non-redundant basis of effective operators for the
  Standard Model Effective Field Theory up to mass dimension~12 with three
  generations of fermions. We also include operators coupling to gravity via
  the Weyl tensor. The results are obtained by implementing the algorithm of
  Li~et~al.~\cite{Li:2020gnx,Li:2020xlh,Li:2022tec}, and provided in the form
  of ancillary files.
\end{abstract}



\section{Introduction}

Owing to the lack of discoveries of particles beyond the \sm, one of the main
research directions over the past decade has been \efts. Given a
renormalizable low-energy theory $\mathcal{L}$ like the \sm,
\efts\ parametrize possible new heavy physics beyond that theory in terms of
operators of mass dimension larger than four. The Wilson coefficients
compensate for the surplus mass dimension through inverse powers of a generic
new-physics scale $\Lambda$ which is assumed to be much larger than the energy
at which this physics is probed.  The operators are composed of the fields of
$\mathcal{L}$ and obey the same gauge symmetries as $\mathcal{L}$.

In order to uniquely constrain the Wilson coefficients by experiment, a number
of possible redundancies among the operators need to be taken into account.
On the one hand, these arise because operators which vanish by \eoms\ can be
eliminated through a re-definition of the field variables in the path
integral\,\cite{Arzt:1993gz,Criado:2018sdb}. Moreover, operators differing by
a total derivative can be identified. Finally, operators which are related by
algebraic identities associated with the group structure of the underlying
internal or Lorentz symmetry can have dependencies.

The \eft\ constructed from the \sm\ is usually referred to as \smeft.  At mass
dimension five, it contains only the lepton-number violating Weinberg operator
(up to generation multiplicities).  Currently, the non-redundant basis of
operators is known up to mass dimension nine\cite{Buchmuller:1985jz,
  Grzadkowski:2010es,Lehman:2014jma,Liao:2016hru,
  Li:2020gnx,Murphy:2020rsh,Liao:2020jmn,Li:2020xlh}.  Beyond that, only the
\textit{number} of independent operators that form a basis\footnote{From now
on, a \textit{basis} shall always denote a complete, non-redundant basis,
unless stated otherwise.} is currently known, albeit broken up into operators
with a specific field content. This number can be obtained from a suitable
Hilbert series\,\cite{Henning:2015alf,Marinissen:2020jmb} or by explicitly
considering the field quantum numbers and permutation
symmetries\,\cite{Fonseca:2017lem}.  The fact that it grows roughly
exponentially with the mass dimension implies that, beyond a certain order in
$1/\Lambda$, it is necessary to give the task of explicitly constructing the
basis to a computer.

In fact, in \citeres{Li:2020gnx,Li:2020xlh}, partly building on concepts
developed in \citere{Fonseca:2019yya}, a
fully algorithmic approach was used to determine the \smeft\ basis up to mass
dimensions eight and nine. Utilizing various functions of
\citeres{Fonseca:2020vke,Fonseca:2011sy},\footnote{Thanks to R.~Fonseca and
the authors of \citere{Li:2022tec} for clarifications on this issue.} its
implementation has been published as a \texttt{Mathematica}
package\,\cite{Li:2022tec}, but its application currently appears to be
restricted to mass dimensions equal to or less than nine.

An automated approach to constructing \eft\ bases is also desirable from the
point of view that as-of-yet undiscovered light particles might still exist
which could couple to the \sm\ via effective operators. Examples for this are
sterile neutrinos~\cite{Li:2021tsq} or axion-like
particles~\cite{Galda:2021hbr}. In addition, one should expect that the
low-energy limit of some theory that includes both the \sm\ interactions as
well as gravity will be described by an effective Lagrangian which extends
\smeft\ by gravitational fields, resulting in
\grsmeft\,\cite{Ruhdorfer:2019qmk}.

Finally, an efficient automated approach to constructing \efts\ will allow one
to study operator bases at higher mass dimension, and thus operators with a
richer structure. For example, it is only starting at mass dimension ten
that $B-L$ can be violated at $\Delta(B-L)=4$, where $B$ and $L$ are the
baryon and lepton number~\cite{Kobach:2016ami,Helset:2019eyc}.
Furthermore, studying higher-dimensional operators, or specific sub-sets
thereof, may give insight on a possible all-order structure of such
operators; see \citere{Jenkins:2009dy}, for example.

In this paper, we report on a re-implementation of the algorithm of
\citere{Li:2020gnx,Li:2020xlh,Li:2022tec} into
\texttt{SageMath}~\cite{sagemath} which is a free open-source mathematics
software system, and the symbolic manipulation system
\texttt{FORM}~\cite{Vermaseren:2000nd,Kuipers:2012rf}, both licensed under the
\gpl~\cite{gpl}.  Using this code, named \texttt{AutoEFT}, we have evaluated
the (on-shell) \smeft\ basis as well as the \grsmeft\ basis up to mass
dimension~12.

A description of \texttt{AutoEFT}, its usage, and the program itself can be
found in \citere{Harlander:2023ozs}. The main purpose of the present
paper is to make the \smeft\ and \grsmeft\ operators of mass dimensions 10,
11, and 12 accessible to the public.\footnote{The lower-dimensional operators
will be provided as well.} However, the number of operators is too large for
listing them in this paper. Instead, we provide them in electronic form, using
a notation which is both rather compact, but also straightforward to
interpret.

The remainder of the paper is structured as follows. In \cref{sec:theory} we
briefly review the algorithm of \citeres{Li:2020gnx,Li:2020xlh,Li:2022tec} on
which the results of this paper are based. This includes the treatment of
scalar fields, chiral fermions, vector bosons, as well as gravitons.
The majority of \cref{sec:smeft} is devoted to describing the format in which we encode
the operators. The results themselves are provided in the form of ancillary
files which accompany this paper. \Cref{sec:conclusions} contains our
conclusions and an outlook.


\section{Constructing EFT Operator Bases}\label{sec:theory}


In this section, the concepts introduced in \citeres{Li:2020gnx,Li:2020xlh}
for the systematic construction of an \eft\ operator basis are briefly
reviewed. Contrary to the formal representation in
\citeres{Li:2020gnx,Li:2020xlh}, only the central ideas are illustrated.


\subsection{Field Representations}\label{sec:theory:fields}

In a standard construction of \smeft\ operators, as it was done for the
Warsaw basis at mass dimension six~\cite{Grzadkowski:2010es}, for example, the
fundamental building blocks are taken to be complex scalars $\phi$, Dirac
spinors $\Psi_{\mathrm{L}/\mathrm{R}}$, field strength tensors $F_{\mu\nu}$,
and the derivative\footnote{Throughout this paper, \textit{derivative} denotes
the gauge covariant derivative, unless stated otherwise.} $D_\mu$. It turns
out that for the operator construction at higher mass dimensions, it is more
convenient to characterize the fields and derivatives by the irreducible
representation $(j_l,j_r)$ in which they transform under the Lorentz group
$\SL(2,\mathbb{C}) \simeq \SU(2)_l \times \SU(2)_r$.  This corresponds to
adopting a chiral basis, where the \sm\ fields are scalars, two-component Weyl
spinors, or chiral field-strength tensors:
\begin{equation}\label[relation]{eq:sl2c}
  \begin{aligned}
    \phi \in &(0,0) \,,\\
    \psi_\alpha \in (1/2,0) \,,&\quad \psi^{\dagger\dot\alpha} \in (0,1/2) \,,\\
    F_{\mathrm{L}\,\alpha\beta} \in (1,0) \,,&\quad
    F_{\mathrm{R}}^{\,\dot\alpha\dot\beta} \in (0,1) \,.
  \end{aligned}
\end{equation}
Here, $\alpha, \beta, \ldots$ and
$\dot\alpha,\dot\beta,\ldots$ denote indices of the fundamental representation
of $\SU(2)_l$ and $\SU(2)_r$, respectively. Note that each of the fields in
\cref{eq:sl2c} has a unique helicity value
\begin{equation}\label{eq:helicity}
  \begin{aligned}
    h=j_r - j_l\,.
  \end{aligned}
\end{equation}
The derivative transforms non-trivially under both $\SU(2)_l$ and $\SU(2)_r$,
\begin{equation}
  \begin{aligned}
    \label[relation]{eq:sl2c_derivative}
    D_\alpha^{\dot\alpha} \in (1/2,1/2) \,.
  \end{aligned}
\end{equation}
Thus, the derivative of a field has the same helicity as the field itself.
This definition of the fields is equivalent to the conventional representation
of the \sm\ fields. The translation between the two notations is given by:
\begin{equation}
    \label{eq:translation}
    \begin{aligned}
      F^{\mu\nu} = \frac{i}{4}\left( F_{\mathrm{L}}^{\,\alpha\beta}
      \sigma^{\mu\nu}_{\alpha\beta}
      -
      F_{\mathrm{R}}^{\,\dot\alpha\dot\beta}\bar{\sigma}^{\mu\nu}_{\dot\alpha\dot\beta}
      \right)\,,
      \quad D_\mu = -\frac{1}{2}D_{\alpha}^{\dot\alpha}
      (\sigma_\mu)^{\alpha}_{\dot\alpha} \\
        \Psi_{\mathrm{L}} =
        \begin{pmatrix}
            \psi_\alpha \\ 0
        \end{pmatrix}
        \,,\quad
        \bar\Psi_{\mathrm{L}} =
        \begin{pmatrix}
            0, \psi^\dagger_{\dot\alpha}
        \end{pmatrix}
        \,,\quad
        \Psi_{\mathrm{R}} =
        \begin{pmatrix}
            0 \\ \psi_{\mathbb{C}}^{\dagger\dot\alpha}
        \end{pmatrix}
        \,,\quad
        \bar\Psi_{\mathrm{R}} =
        \begin{pmatrix}
            \psi_{\mathbb{C}}^\alpha, 0
        \end{pmatrix}
        \,,\\
        F_{\mathrm{L}\,\alpha\beta} = \frac{i}{2}F_{\mu\nu}\sigma^{\mu\nu}_{\alpha\beta}
        \,,\quad
        F_{\mathrm{R}}^{\,\dot\alpha\dot\beta} = -\frac{i}{2}F^{\mu\nu}\bar{\sigma}_{\mu\nu}^{\dot\alpha\dot\beta}
        \,,\quad
        D_\alpha^{\dot\alpha} = D_\mu (\sigma^\mu)_\alpha^{\dot\alpha}
        \,,
    \end{aligned}
\end{equation}
where $\psi_{\mathbb{C}}$ denotes a charge conjugated spinor,
and the $\sigma$-matrices are given by
\begin{equation}
    \label{eq:sigma}
    \begin{aligned}
        \sigma^{\mu\nu} =
        \frac{i}{2}\left(\sigma^\mu\bar\sigma^\nu-\sigma^\nu\bar\sigma^\mu\right)
        \,,&\quad
        \bar{\sigma}_{\mu\nu} =
        \frac{i}{2}\left(\bar\sigma_\mu\sigma_\nu-\bar\sigma_\nu\sigma_\mu\right)
        \,,\\
        \sigma^\mu_{\alpha\dot\alpha} = (I,\vec{\sigma})
        \,,&\quad
        \bar\sigma^{\mu\,\dot\alpha\alpha} = (I,-\vec{\sigma})
        \,.
    \end{aligned}
\end{equation}
Here, $I$ is the $2 \times 2$ identity matrix and $\vec{\sigma} =
\left(\sigma^1,\sigma^2,\sigma^3\right)$ denotes the Pauli matrices.


\subsection{Construction of the Operators}\label{sec:construct}

The operators can be classified into \textit{families}\footnote{In
\citere{Li:2020gnx} the term \textit{subclass} is used instead.},
characterized by the tuple
\begin{equation}\label{eq:tuple}
  \begin{aligned}
        (n_{F_{\mathrm{L}}}, n_{\psi}, n_{\phi}, n_{\psi^\dagger}, n_{F_{\mathrm{R}}}; n_D)\,,
  \end{aligned}
\end{equation}
where $n_\Phi$ equals the number of fields with helicity $h_\Phi$, and $n_D$
is the number of derivatives. It is further useful to define
\begin{equation}
  \begin{aligned}
        \label{eq:nl_nr}
    n_l = n_{F_{\mathrm{L}}} + \frac12 n_{\psi} + \frac12 n_D \,,\qquad n_r =
    n_{F_{\mathrm{R}}} + \frac12 n_{\psi^\dagger} + \frac12 n_D\,,
  \end{aligned}
\end{equation}
which correspond to the sum of $j_{l/r}$ of each field and derivative as
defined in \cref{eq:sl2c}.  Lorentz invariance and a given mass dimension $d$
of the operators put constraints on the $n_i$ such as
\begin{equation}\label{eq:goya}
  \begin{aligned}
    N+n_l+n_r=d\,,
  \end{aligned}
\end{equation}
where $N$ is the total number of fields, and thus restrict the set of allowed
families.

Since Hermitian conjugation exchanges the two $\SU(2)$ representations of the
Lorentz group, taking the conjugate of the operators of a particular family
generates the \textit{conjugate family}
\begin{equation}
    (n_{F_{\mathrm{L}}}, n_{\psi}, n_{\phi}, n_{\psi^\dagger},
  n_{F_{\mathrm{R}}}; n_D)^\dagger \equiv (n_{F_{\mathrm{R}}},
  n_{\psi^\dagger}, n_{\phi}, n_{\psi}, n_{F_{\mathrm{L}}}; n_D) \,.
\end{equation}
Hence, one can identify \textit{real} families that satisfy
\begin{equation}
    (n_{F_{\mathrm{L}}}, n_{\psi}, n_{\phi}, n_{\psi^\dagger}, n_{F_{\mathrm{R}}}; n_D)^\dagger = (n_{F_{\mathrm{L}}}, n_{\psi}, n_{\phi}, n_{\psi^\dagger}, n_{F_{\mathrm{R}}}; n_D) \,.
\end{equation}
All operators in a real family are either Hermitian, or their conjugate operator
is part of the same family.  The remaining families are all \textit{complex},
such that any operator features a distinct Hermitian conjugate version, which
is part of the conjugate family.

All operators in a specific family can be further characterized by their
\textit{type}, which corresponds to a specific multiset of fields from a given
model.  Types of the same field content are identified by ordering the fields
by increasing helicity and sorting fields of the same helicity
alphanumerically.  For example, the dimension-five Weinberg operator,
consisting of two lepton doublets and two Higgs doublets, belongs to the
family $(0,2,2,0,0;0)$ and the type $L^2H^2$.


\subsection{Lorentz Structure}\label{sec:theory:lorentz}

In this section, only the Lorentz structure of an operator will be considered,
while all internal symmetry and generation indices of the fields will be
neglected. All operators of a particular family can thus be identified for the
purpose of this discussion.

Since any operator $\mathcal{O}$ of the \eft\ will be constructed from the
objects of \cref{eq:sl2c,eq:sl2c_derivative}, it will be of the form
\begin{equation}\label{eq:lorentz}
    \mathcal{O}^{\text{Lorentz}} =
    (T^{\text{Lorentz}})_{\boldsymbol{\dot\alpha}_1
      \ldots\boldsymbol{\dot\alpha}_N}^{\boldsymbol{\alpha}_1\ldots\boldsymbol{\alpha}_N}
    \prod_{i=1}^N
    (D^{n_i}\Phi_i)_{\boldsymbol{\alpha}_i}^{\boldsymbol{\dot\alpha}_i} \,,
\end{equation}
where $\boldsymbol{\alpha}_i = (\alpha_i^{(1)},\dots,\alpha_i^{(m_i)})$ and
$\boldsymbol{\dot\alpha}_i = (\dot\alpha_i^{(1)},\dots,\dot\alpha_i^{(\tilde{m}_i)})$ are multi-indices with $m_i = n_i + 2j_{l,i}$ and $\tilde{m}_i = n_i
+ 2j_{r,i}$, where $n_i$ is the number of derivatives acting on the field
$\Phi_i$, and $(j_{l,i},j_{r,i})$ defines its representation according to
\cref{eq:sl2c}. The indices $\alpha_i^{(1)},\ldots
\alpha_i^{(n_i)}$ and $\dot\alpha^{(1)}_i,\ldots,\dot\alpha^{(n_i)}_i$ are to be
associated with the derivatives acting on $\Phi_i$, while the remaining
indices are associated with the field itself.


\subsection{Redundancies}\label{sec:theory:redundancies}

One of the most challenging aspects when constructing \efts\
is the elimination of redundancies, i.e.~operators that are related to other
operators by certain identities. As stated in \citere{Li:2020gnx},
for the chiral convention of the fields, 
the structure of $T^{\text{Lorentz}}$ must be a polynomial in
$\epsilon^{\alpha\beta}$ and $\epsilon_{\dot\alpha\dot\beta}$, where
$\epsilon$ is totally anti-symmetric, and $\epsilon^{12} = \epsilon_{21} =
\epsilon^{\dot1\dot2} = \epsilon_{\dot2\dot1} = 1$. This already eliminates
all redundancies from Fierz identities (up to Schouten
identities). Furthermore, redundancies due to \eoms\ or the identity
$i[D_\mu,D_\nu] = \sum F_{\mu\nu}$, where the sum is over all gauge groups,
can be eliminated by substituting
\begin{equation}\label[relation]{eq:edom}
  \begin{aligned}
    (D^{n_i}\Phi_i)_{\boldsymbol{\alpha}_i}^{\boldsymbol{\dot\alpha}_i}
    \to 
    (D^{n_i}\Phi_i)_{(\boldsymbol{\alpha}_i)}^{(\boldsymbol{\dot\alpha}_i)}
  \end{aligned}
\end{equation}
in \cref{eq:lorentz}, where $(\boldsymbol{\alpha}_i) =
(\alpha_i^{(1)},\dots,\alpha_i^{(m_i)})_{\text{sym}}$ and
$(\boldsymbol{\dot\alpha}_i) = (\dot\alpha_i^{(1)},\dots,\dot\alpha_i^{(\tilde{m}_i)})_{\text{sym}}$ denote totally symmetric multi-indices.  In the
following, we refer to the combined object on the r.h.s.\ of \cref{eq:edom}
as \textit{building block}.

The remaining redundancies to be considered are due to \ibp\ and Schouten
identities. These can be avoided by introducing an auxiliary $\SU(N)$ group,
where $N$ equals the total number of fields (cf.~\cref{eq:goya}).  Under this
group, the (dotted) spinor indices of $T^{\text{Lorentz}}$ transform as the
(anti-)fundamental representation, i.e.\
\begin{equation}\label{eq:theory:kula}
  \begin{aligned}
    \boldsymbol{\alpha}_i\to \sum_{j=1}^N U_{ij}\boldsymbol{\alpha}_j\,,\qquad
    \boldsymbol{\dot\alpha}_i\to \sum_{j=1}^N
    U^\dagger_{ij}\boldsymbol{\dot\alpha}_j\,,
  \end{aligned}
\end{equation}
with $\boldsymbol{\alpha}_i$, $\boldsymbol{\dot{\alpha}}_i$ defined in
\cref{eq:lorentz}, and $U$ ($U^\dagger$) group elements of $SU(N)$ in the
\mbox{(anti-)}\-fun\-da\-men\-tal representation. For a given
family, $T^{\text{Lorentz}}$ contains $n_l$ $\epsilon$-tensors with undotted
and $n_r$ $\epsilon$-tensors with dotted indices.  Consequently,
$T^{\text{Lorentz}}$ must transform in the representation
\begin{equation}
  \mathcal{R} =
  \overline{\ydiagram{1,1}}^{\,\otimes n_r}
  \otimes
  \quad
  \ydiagram{1,1}^{\,\otimes n_l}
\end{equation}
under the auxiliary $SU(N)$ group.
According to the \lwrs~rule, the two factors can be decomposed as
\begin{equation}\label{eq:schouten}
  \mathcal{R} =
  \bigg(\overbrace{\overline{
    \begin{ytableau}
        ~ & \ytabdots & ~ \\
        ~ & \ytabdots & ~
    \end{ytableau}
    }
    \vphantom{\ydiagram{1,1,1}}
    }^{n_r}
    \quad\oplus\quad\dots\bigg)\otimes\bigg(
    \overbrace{
    \begin{ytableau}
        ~ & \ytabdots & ~ \\
        ~ & \ytabdots & ~
    \end{ytableau}
    \vphantom{\ydiagram{1,1,1}}
    }^{n_l}
    \quad\oplus\quad\dots\bigg)\,,
\end{equation}
where dropping all diagrams represented by the dots eliminates the Schouten
identities.  The further decomposition of \cref{eq:schouten} into
irreducible representations results in a Young diagram of shape
\begin{equation}\label{eq:lambda}
    \lambda \quad=\quad
    \rotatebox[origin=c]{90}{$\scriptstyle N-2$}
    \left\{\vphantom{\ydiagram{1,1,1,1}}\right.
    \overbrace{
    \begin{ytableau}
        ~ & \ytabdots & ~ \\
        ~ & \ytabdots & ~ \\
        \ytabvdots & \none & \ytabvdots \\
        ~ & \ytabdots & ~
    \end{ytableau}
    \vphantom{\ydiagram{1,1,1,1,1}}
    }^{n_r}\!
    \overbrace{
    \begin{ytableau}
        ~ & \ytabdots & ~ \\
        ~ & \ytabdots & ~ \\
        \none \\
        \none
    \end{ytableau}
    \vphantom{\ydiagram{1,1,1,1,1}}
    }^{n_l}
	\quad,
\end{equation}
plus diagrams that contain at least one column with $N-1$ entries which will
be discussed below. A basis of tensors that transform under this irreducible
representation can be constructed from \ssytx\ of shape $\lambda$ and
content\footnote{A Young tableau with content (or weight) $\mu = [i, j, k,
  \dots]$ has $i$ entries of the number $1$, $j$ entries of the number $2$,
$k$ entries of the number $3$, and so on.  This requirement ensures that the
resulting tensors have the correct indices corresponding to the fields
specified in the family. A Young tableau is called semi-standard if the
entries weakly increase along each row and strictly increase down each
column.}  $\mu = [n_r-2h_1, \dots, n_r-2h_N]$
(cf.~\cref{eq:nl_nr,eq:helicity,eq:goya}) as follows: Replace the
first $n_r$ columns through the $\SU(N)$ relation (no summation implied)
\begin{equation}
\label{eq:conj_epsilon}
    \begin{ytableau}
        a \\
        b \\
        \ytabvdots \\
        x
    \end{ytableau}
    \quad=\quad
    \mathcal{E}^{ab \dots xyz}\ 
    \overline{\ytableaushort{{y},{z}}} \,,
\end{equation}
where $\mathcal{E}$ denotes the $N$-dimensional Levi-Civita symbol, and
$a,b,\dots,x,y,z$ is some permutation of $1,2,\dots,N$.  Subsequently
identify each column with an $\epsilon$-tensor according to
\begin{equation}\label{eq:epsun}
    \overline{\ytableaushort{i,j}} \sim \epsilon_{\dot\alpha_i\dot\alpha_j}
    \qquad\text{and}\qquad \ytableaushort{i,j} \sim
    \epsilon^{\alpha_i\alpha_j}
	\,.
\end{equation}
Since the set of all \ssytx\ forms a basis in the vector space of a particular
representation, performing these steps for every such \ssyt\ leads to a set of
Lorentz tensors $T^\text{Lorentz}$ which is non-redundant.  To see that it is
also complete, consider a Young diagram other than \cref{eq:lambda} in the
decomposition of \cref{eq:schouten}. As pointed out above, it contains at
least one column of length $N-1$. Such columns correspond to tensors
proportional to $\sum_{j=1}^N
\epsilon^{\alpha_i\alpha_j}\epsilon_{\dot\alpha_j\dot\alpha_k}$. The index
pair ($\alpha_j,\dot{\alpha}_j$) implies a derivative acting on the $j$-th
field, so that the sum corresponds to a total derivative.


\subsection{Internal Symmetries}\label{sec:theory:gauge}

After identifying all independent Lorentz structures, one can continue in a
similar manner with the remaining symmetries of the underlying low-energy
theory.  In the following, the expression \textit{internal symmetry} refers to a
global or a local $\U(1)$ or $\SU(n)$ symmetry. In analogy to
\cref{sec:theory:lorentz}, all Lorentz and generation indices will be
neglected in this section, and only the transformation properties of the
fields under the internal symmetry group are considered.

Concerning Abelian internal symmetries, it is required that the total charge
of an operator under this symmetry vanishes.  In the \sm, for example, this is
achieved by only considering combinations of fields such that the sum of their
hypercharges equals zero, hence forming a $\U(1)$ invariant operator. However,
one may also allow for the breaking of a certain $\U(1)$ symmetry, specified
by the amount the associated charge of the operators may deviate from zero.
This can be useful, for example, if one is only interested in operators that
violate baryon number conservation up to a certain degree $\Delta B$.

For each non-Abelian symmetry, the \textit{modified \lwrs~rule}~\cite{Li:2020gnx}
is used to construct all sets of independent tensors.  To
apply this method, the fields of the low-energy theory must be characterized
in terms of fundamental $\SU(n)$ indices only. Consider, for example, the
non-Abelian part of the \sm, i.e.,~$\SU(3)\otimes\SU(2)$. In terms of Young
diagrams, the fundamental representations are identified as $\boldsymbol{3} =
\ydiagram{1}$ for $\SU(3)$, and $\boldsymbol{2} = \ydiagram{1}$ for $\SU(2)$.
For the \sm\ fields that transform under the fundamental representations, the
symmetry of their indices can be identified with the states
\begin{equation}
    Q_{a\,i} \sim \ytableaushort{a}\otimes\ytableaushort{i} \,,\qquad q_a \sim
    \ytableaushort{a} \,,\qquad L_i \sim \ytableaushort{i}
    \,,\quad\text{and}\qquad H_i \sim \ytableaushort{i} \,,
\end{equation}
where $Q$ and $L$ denote a left-handed quark and lepton doublet, respectively;
$q$ the right-handed quark fields; and $H$ the Higgs doublet. Here,
$a,b,c,\ldots$ and $i,j,k,\ldots$ denote fundamental indices of $\SU(3)$ and
$\SU(2)$, respectively.

The fields that do not transform under the fundamental representation of the
internal symmetry group can nevertheless be written in terms of quantities
with fundamental indices only. In particular, for the \sm,
\begin{equation}\label{eq:cube}
  \begin{aligned}
    G_{abc} &= \epsilon_{acd}(\lambda^A)\indices{^d_b} G^A \,,& \qquad W_{ij} &=
    \epsilon_{jk}(\tau^I)\indices{^k_i}W^I \,,\\ Q^\dagger_{ab\,i} &=
    \epsilon_{abc}\epsilon_{ij}(Q^\dagger)^{c\,j} \,,&\qquad q^\dagger_{ab} &=
    \epsilon_{abc}(q^\dagger)^c \,,\\ L^\dagger_i &=
    \epsilon_{ij}(L^\dagger)^j \,,&\qquad H^\dagger_i &=
    \epsilon_{ij}(H^\dagger)^j \,,
  \end{aligned}
\end{equation}
where $A$ and $I$ are adjoint indices of $\SU(3)$ and $\SU(2)$.  The
$\lambda^A$ are the Gell-Mann matrices for $\SU(3)$, and the $\tau^I$ are
the $\SU(2)$ Pauli matrices.

Considering the Young diagrams for the adjoint representation of $\SU(3)$ and
$\SU(2)$, given by $\boldsymbol{8} = \ydiagram{2,1}$ and $\boldsymbol{3} =
\ydiagram{2}$, one can identify the symmetry of the fundamental symmetry group
indices of $G_{abc}$ and $W_{ij}$ with the states
\begin{equation}
    G_{abc} \sim \ytableaushort{ab,c} \qquad\text{and}\qquad W_{ij} \sim \ytableaushort{ij} \,,
\end{equation}
meaning that $G_{abc}$ is symmetrized in $a \leftrightarrow b$ and
subsequently anti-symmetrized in $a \leftrightarrow c$, while $W_{ij}$ is
totally symmetric in $i \leftrightarrow j$.  Equivalently, the
anti-fundamental representations of $\SU(3)$ and $\SU(2)$ are given by
$\boldsymbol{\bar3} = \overline{\ydiagram{1}} = \ydiagram{1,1}$ and
$\boldsymbol{\bar 2} = \overline{\ydiagram{1}} = \ydiagram{1}$, respectively.
The symmetry of the fundamental symmetry group
indices can be identified with
\begin{equation}
    Q^\dagger_{ab\,i} \sim \ytableaushort{a,b}\otimes\ytableaushort{i} \,,\qquad q^\dagger_{ab} \sim \ytableaushort{a,b} \,,\qquad L^\dagger_i \sim \ytableaushort{i} \,,\quad\text{and}\qquad H^\dagger_i \sim \ytableaushort{i} \,,
\end{equation}
where the only non-trivial symmetry is given by $Q^\dagger_{ab\,i}$ and $q^\dagger_{ab}$ which are anti-symmetric under the exchange $a \leftrightarrow b$.

From the
discussion above, it is clear that one can write a generic operator as
\begin{equation}\label{eq:gauge}
    \mathcal{O}^{\SU(n)} = (T^{\SU(n)})^{\boldsymbol{I}_1\ldots\boldsymbol{I}_N}
    \prod_{i=1}^N (\Phi_i)_{\boldsymbol{I}_i} \,,
\end{equation}
where $\boldsymbol{I}_i$ is a multi-index containing $\SU(n)$ fundamental
indices only.  The independent set of tensors $T^{\SU(n)}$ is constructed by
combining the Young tableaux of each field in the operator according to the
modified \lwrs~rule, and subsequently identifying each column of the resulting
tableau with an $\SU(n)$ $\epsilon$-tensor.


\subsection{Permutation Symmetries}\label{sec:theory:permutation}
It is important to note that, up to this point, the algorithm considers all
fields that occur in an operator as distinct. But if an operator contains
several copies of the same field, new redundancies can occur. This happens
because in expressions like \cref{eq:lorentz,eq:gauge}, the sum over indices
is not explicitly carried out. Therefore, the Lorentz and internal symmetry
groups can no longer be treated independently as soon as the operator contains
identical fields.

As a generalization of this, a theory may contain several copies of fields
that transform identically under the Lorentz and internal symmetry groups
and are thus indistinguishable for our considerations.
In the spirit of the \sm, we will refer to these copies as
\textit{generations}\footnote{\citere{Li:2020gnx} uses the term \textit{flavors}
instead.}.  Owing to the Lorentz and internal symmetry of the operator, not all
combinations of generations are independent of one another, in general.

Both of these problems can be treated in the same way when introducing
\textit{generation indices} for \textit{all} fields, even for those that occur only
in a single generation.  The expression \textit{repeated fields} denotes the
product of fields which at most differ by their generation index.  The general
strategy is to decompose any operator into a sum of terms with specific
permutation symmetry $\lambda$ of the generation indices for repeated
fields~\cite{Fonseca:2019yya}.\footnote{In general, the symbol $\lambda$
can denote multiple irreducible representations of the symmetric group, one for each set of repeated
fields.}

Following the philosophy of \citere{Li:2020gnx}, the permutation
symmetry is not inscribed on the operator by symmetrizing the generation
indices themselves, but at the level of the Lorentz and internal symmetry
tensors they are multiplied with. Consider a specific \textit{type}, i.e.~the
set of operators which contain a certain set of fields. Up to now, the
algorithm has generated a set of tensors
$T^{\text{Lorentz}}=\{T^{\text{Lorentz}}_1\,,\ldots,\,T^{\text{Lorentz}}_l\}$
for the Lorentz symmetry, and
$T^{\SU(n_k)}=\{T^{\SU(n_k)}_1\,,\ldots,\,T^{\SU(n_k)}_{l_k}\}$ for each internal
symmetry; see \cref{sec:theory:lorentz,sec:theory:gauge}.  These
tensors are now combined in such a way that they reflect the permutation
symmetries, labeled by $\lambda$:
\begin{equation}
\label{eq:permutation}
\begin{split}
    \mathcal{T}^\lambda_j &= \sum_i \mathcal{K}^\lambda_{ji} \left(
    T^{\text{Lorentz}} \otimes T^{\SU(n_1)} \otimes \cdots \otimes
    T^{\SU(n_k)} \right)_i \,,
\end{split}
\end{equation}
where the $\mathcal{K}$ are obtained from the plethysm technique and
inner-product decomposition (for details, see \citere{Li:2020gnx}). The
redundancies due to the permutation symmetries are reflected in the fact that
$\mathcal{K}$ is an $n\times m$ matrix with $n\leq m = l\cdot l_1 \cdots l_k$
in general.  The set of independent operators is then given by contracting the
fields with $\mathcal{T}^\lambda_j$ for each $j=1,\dots,n$ and each permutation symmetry $\lambda$,
and enumerating the generation indices according to the set of \ssytx.
Explicit examples will be given below.



\subsection{Including Gravity}\label{sec:gravity}

The algorithm described above can be extended to theories which include fields
with higher spin\,\cite{Ruhdorfer:2019qmk,Durieux:2019siw,Li:2022tec}. In
particular, gravitational interactions can be taken into account by
considering the Weyl tensor $C_{\mu\nu\rho\sigma}$ as an additional building
block in the \eft\ construction~\cite{Ruhdorfer:2019qmk}:
\begin{equation}
    \label{eq:weyl_tensor}
    C_{\mu\nu\rho\sigma} = R_{\mu\nu\rho\sigma} - \left( g_{\mu [ \rho}R_{\sigma ] \nu} - g_{\nu [ \rho}R_{\sigma ] \mu} \right) + \frac13 g_{\mu [ \rho}g_{\sigma ] \nu}R \,,
\end{equation}
where $R_{\mu\nu\rho\sigma}$ denotes the Riemann tensor,
$R_{\mu\nu}=R\indices{^\lambda_\mu_\lambda_\nu}$ the Ricci tensor,
$R=R\indices{^\mu_\mu}$ the Ricci scalar, and $g_{\mu\nu}$ the metric tensor, and the indices between square brackets are to be anti-symmetrized.
In analogy to \cref{sec:theory:fields}, we write the Weyl tensor in terms of
left- and right-handed components that transform under irreducible
representations $(j_l,j_r)$ of the Lorentz group:
\begin{equation}\label{eq:gravity:eats}
  \begin{aligned}
    C^{\mu\nu\rho\sigma} &= \frac{1}{64} \left(
    C_\mathrm{L}^{\,\alpha\beta\gamma\delta}
    \sigma^{\mu\nu}_{\alpha\beta}\sigma^{\rho\sigma}_{\gamma\delta}
    + 
    C_{\mathrm{R}}^{\,\dot\alpha\dot\beta\dot\gamma\dot\delta}
      \bar{\sigma}^{\mu\nu}_{\dot\alpha\dot\beta}
      \bar{\sigma}^{\rho\sigma}_{\dot\gamma\dot\delta} \right)\,,
  \end{aligned}
\end{equation}
where
\begin{equation}\label{eq:sl2c:gr}
  \begin{aligned}
    C_{\mathrm{L}\,\alpha\beta\gamma\delta}
    = C_{\mu\nu\rho\sigma}\sigma^{\mu\nu}_{\alpha\beta}\sigma^{\rho\sigma}_{\gamma\delta}
    \in (2,0) \,,
      \quad\text{and}\quad
      C_{\mathrm{R}}^{\,\dot\alpha\dot\beta\dot\gamma\dot\delta}
    = C^{\mu\nu\rho\sigma}\bar{\sigma}_{\mu\nu}^{\dot\alpha\dot\beta}\bar{\sigma}_{\rho\sigma}^{\dot\gamma\dot\delta}
      \in (0,2) \,.
  \end{aligned}
\end{equation}
Denoting by $n_{C_\mathrm{L/R}}$ the number of left-/right-handed Weyl tensors
in the operators, we extend the families \labelcref{eq:tuple} to
\begin{equation}
    (n_{C_{\mathrm{L}}}, n_{F_{\mathrm{L}}}, n_{\psi}, n_{\phi}, n_{\psi^\dagger}, n_{F_{\mathrm{R}}}, n_{C_{\mathrm{R}}}; n_D)\,.
\end{equation}

While all previously considered fields satisfy the relation
\begin{equation}
    [\Phi] = 1 + \lvert h_\Phi \rvert \,,
\end{equation}
with $[\cdot]$ denoting the mass dimension in four space-time dimensions and
$h_\Phi$ the helicity, this relation is violated for
the Weyl tensor, which satisfies
\begin{equation}
    \label{eq:gr_massdim}
    [C_{\mathrm{L}/\mathrm{R}}] = \lvert h_{C_{\mathrm{L}/\mathrm{R}}} \rvert = 2 \,,
\end{equation}
instead. Therefore, the algorithm needs to account for the actual mass dimension of the Weyl
tensor. For example, instead of \cref{eq:goya}, we get
\begin{equation}\label{eq:gravity:evan}
  \begin{aligned}
    N+ n_l + n_r - n_{\mathrm{GR}} = d \,,
  \end{aligned}
\end{equation}
where $n_{\mathrm{GR}} \equiv n_{C_{\mathrm{L}}} + n_{C_{\mathrm{R}}}$ can
take values from $0,\dots,\min(d/2, N)$, and the definitions
in \cref{eq:nl_nr} are modified, such that
\begin{equation}
    \begin{aligned}
        \label{eq:nl_nr:gr}
    n_l = 2n_{C_{\mathrm{L}}} + n_{F_{\mathrm{L}}} + \frac12 n_{\psi} + \frac12 n_D \,,\qquad n_r =
    2n_{C_{\mathrm{R}}} + n_{F_{\mathrm{R}}} + \frac12 n_{\psi^\dagger} + \frac12 n_D\,.
  \end{aligned}
\end{equation}
An example for an operator type of dimension-12 in
\grsmeft\ will be given below.


\section{Application to SMEFT and GRSMEFT}\label{sec:smeft}


According to the algorithm described in \cref{sec:theory}, we construct
on-shell operator bases for \smeft\ and \grsmeft\ up to mass dimension 12,
using our implementation \texttt{AutoEFT}. To be consistent with the existing
literature, we adopt the all-left chirality convention for the
\sm\ fields. The \eft\ field content is given in \cref{tab:sm_fields}.  These
fields are related to the conventional notation by
\cref{eq:translation,eq:cube,eq:gravity:eats,eq:sl2c:gr}.  For example, the
\sm\ Dirac spinors are given by
\begin{equation}
\begin{aligned}
    Q_{\mathrm{L}} =
    \begin{pmatrix}
        Q_\alpha \\ 0
    \end{pmatrix}
    \,,\quad
    u_{\mathrm{R}} &=
    \begin{pmatrix}
        0 \\ u_{\mathbb{C}}^{\dagger\dot\alpha}
    \end{pmatrix}
    \,,\quad
    d_{\mathrm{R}} =
    \begin{pmatrix}
        0 \\ d_{\mathbb{C}}^{\dagger\dot\alpha}
    \end{pmatrix}
    \,,\\[1em]
    L_{\mathrm{L}} =
    \begin{pmatrix}
        L_\alpha \\ 0
    \end{pmatrix}
    \,&,\quad
    e_{\mathrm{R}} =
    \begin{pmatrix}
        0 \\ e_{\mathbb{C}}^{\dagger\dot\alpha}
    \end{pmatrix}
    \,,
\end{aligned}
\end{equation}
where all internal symmetry and generation indices are suppressed.


\begin{table}
    \centering
    \renewcommand*{\arraystretch}{1.55}
    \begin{tabular}{crllrcc}
        \toprule
        Field                                     & $h$    & $\SU(3)$              & $\SU(2)$           & $\U(1)$ & generations & \texttt{symbol} \\
        \midrule
        $(G_{\mathrm{L}\,\alpha\beta})_{abc}$     & $-1$   & \ytableaushort{ab,c}  & -                  & $0$     & $1$         & \texttt{GL}     \\
        $(W_{\mathrm{L}\,\alpha\beta})_{ij}$      & $-1$   & -                     & \ytableaushort{ij} & $0$     & $1$         & \texttt{WL}     \\
        $B_{\mathrm{L}\,\alpha\beta}$             & $-1$   & -                     & -                  & $0$     & $1$         & \texttt{BL}     \\
        $(Q_\alpha^{g})_{a\,i}$                   & $-1/2$ & \ytableaushort{a}     & \ytableaushort{i}  & $1/6$   & $3$         & \texttt{Q}      \\
        $(u_{\mathbb{C}\,\alpha}^g)_{ab}$         & $-1/2$ & \ytableaushort{a,b}   & -                  & $-2/3$  & $3$         & \texttt{uC}     \\
        $(d_{\mathbb{C}\,\alpha}^g)_{ab}$         & $-1/2$ & \ytableaushort{a,b}   & -                  & $1/3$   & $3$         & \texttt{dC}     \\
        $(L_\alpha^g)_{i}$                        & $-1/2$ & -                     & \ytableaushort{i}  & $-1/2$  & $3$         & \texttt{L}      \\
        $e_{\mathbb{C}\,\alpha}^g$                & $-1/2$ & -                     & -                  & $1$     & $3$         & \texttt{eC}     \\
        $H_i$                                     & $0$    & -                     & \ytableaushort{i}  & $1/2$   & $1$         & \texttt{H}      \\
        \midrule
        $C_{\mathrm{L}\,\alpha\beta\gamma\delta}$ & $-2$   & -                     & -                  & $0$     & $1$         & \texttt{CL}     \\
        \bottomrule
    \end{tabular}
    \caption{\sm\ field content in the all-left chiral convention
      (cf.~\cref{sec:theory:fields}).  Spinor indices are denoted by $\alpha$
      and $\beta$, while $a,b,c$ and $i,j$ denote fundamental indices of the
      gauge groups $\SU(3)$ and $\SU(2)$, respectively.  Each fermion carries
      an additional generation index $g$.  The Young tableaux of the $\SU(n)$
      groups shown in columns three and four have been introduced in
      \cref{sec:theory:gauge}.  The \texttt{symbol} denotes the characters
      that represent the corresponding field in the ancillary files.
      Hermitian conjugated fields are denoted by a trailing ``\texttt{+}''.
      The covariant derivative is represented by the character ``\texttt{D}''.
    }\label{tab:sm_fields}
\end{table}


For a given mass dimension, the operators are grouped into families
\labelcref{eq:tuple}, and in each family, the operators are classified by
their type (i.e.,~the field content).  For each of these types, an output file
is generated in \texttt{YAML} format.\footnote{See \url{https://yaml.org/}.}
It contains the relevant information to construct all operators belonging to
this type. We found this way of presenting the results a reasonable compromise
which allows one to encode the large number of operators in a rather compact form
without leaving too much calculational effort to the user.

In the following, four representative cases are displayed to show
how the output files can be interpreted in terms of \cref{sec:theory}.


\subsection{No Repeated Fields}\label{sec:norepeat}

As a first example, let us consider the dimension-10 operators of type
\begin{equation}\label{eq:kush}
  \begin{aligned}
      \mathcal{Q}_\text{NRF} &= W_{\mathrm{L}}Qu_\mathbb{C}HW^\dagger_{\mathrm{L}} D^2\,,
  \end{aligned}
\end{equation}
encoded in the file
\begin{verbatim}
  operators/5/1FL_2psi_1phi_1FR_2D/1WL_1Q_1uC_1H_1WL+_2D.yml
\end{verbatim}
included in \verb$smeft_10.tar.xz$. They contain a weak gauge field and its
conjugate (\verb$1WL$, \verb$1WL+$), a quark doublet (\verb$1Q$), a
charge-conjugate up-type quark singlet (\verb$1uC$), a Higgs field
(\verb$1H$), and two derivatives (\verb$2D$). This also illustrates the
naming scheme of the output files of \texttt{AutoEFT}. The content of this file
is displayed in \cref{lst:dim10}. We deliberately include redundant
information in the output files in order to facilitate their interpretation
and to allow for consistency checks when constructing the explicit operators.

\begin{lstlisting}[caption={Content of the file
    \texttt{1WL\symbol{95}1Q\symbol{95}1uC\symbol{95}1H\symbol{95}1W+\symbol{95}2D.yml},
    which encodes the operators of \cref{eq:kush}.}, label={lst:dim10},
  style=yaml]
# '1WL_1Q_1uC_1H_1WL+_2D.yml' generated by AutoEFT 1.0.0
version: 1.0.0
type:
- {WL: 1, Q: 1, uC: 1, H: 1, WL+: 1, D: 2}
- complex
generations: {WL: 1, Q: 3, uC: 3, H: 1, WL+: 1}
n_terms: 10
n_operators: 90
invariants:
  Lorentz:
    O(Lorentz,1): +eps(1_1,3_1)*eps(1_2,3_2)*eps(2_1,3_3)*eps(3_1~,5_1~)*eps(3_2~,5_2~) * WL(1_1,1_2)*Q(2_1)*(D^2 uC)(3_1,3_2,3_3,3_1~,3_2~)*H*WL+(5_1~,5_2~)
    O(Lorentz,2): -eps(1_1,3_1)*eps(1_2,3_2)*eps(2_1,4_1)*eps(3_1~,5_1~)*eps(4_1~,5_2~) * WL(1_1,1_2)*Q(2_1)*(D uC)(3_1,3_2,3_1~)*(D H)(4_1,4_1~)*WL+(5_1~,5_2~)
    O(Lorentz,3): +eps(1_1,3_1)*eps(1_2,4_1)*eps(2_1,4_2)*eps(4_1~,5_1~)*eps(4_2~,5_2~) * WL(1_1,1_2)*Q(2_1)*uC(3_1)*(D^2 H)(4_1,4_2,4_1~,4_2~)*WL+(5_1~,5_2~)
    O(Lorentz,4): -eps(1_1,2_1)*eps(1_2,3_1)*eps(3_2,4_1)*eps(3_1~,5_1~)*eps(4_1~,5_2~) * WL(1_1,1_2)*Q(2_1)*(D uC)(3_1,3_2,3_1~)*(D H)(4_1,4_1~)*WL+(5_1~,5_2~)
    O(Lorentz,5): +eps(1_1,2_1)*eps(1_2,4_1)*eps(3_1,4_2)*eps(4_1~,5_1~)*eps(4_2~,5_2~) * WL(1_1,1_2)*Q(2_1)*uC(3_1)*(D^2 H)(4_1,4_2,4_1~,4_2~)*WL+(5_1~,5_2~)
  SU3:
    O(SU3,1): +eps(2_1,3_1,3_2) * WL*Q(2_1)*uC(3_1,3_2)*H*WL+
  SU2:
    O(SU2,1): +eps(1_1,4_1)*eps(1_2,5_1)*eps(2_1,5_2) * WL(1_1,1_2)*Q(2_1)*uC*H(4_1)*WL+(5_1,5_2)
    O(SU2,2): +eps(1_1,2_1)*eps(1_2,5_1)*eps(4_1,5_2) * WL(1_1,1_2)*Q(2_1)*uC*H(4_1)*WL+(5_1,5_2)
permutation_symmetries:
- vector: Lorentz * SU3 * SU2
- symmetry: {WL: [1], Q: [1], uC: [1], H: [1], WL+: [1]}
  n_terms: 10
  n_operators: 90
  matrix: |-
    [1 0 0 0 0 0 0 0 0 0]
    [0 1 0 0 0 0 0 0 0 0]
    [0 0 1 0 0 0 0 0 0 0]
    [0 0 0 1 0 0 0 0 0 0]
    [0 0 0 0 1 0 0 0 0 0]
    [0 0 0 0 0 1 0 0 0 0]
    [0 0 0 0 0 0 1 0 0 0]
    [0 0 0 0 0 0 0 1 0 0]
    [0 0 0 0 0 0 0 0 1 0]
    [0 0 0 0 0 0 0 0 0 1]
\end{lstlisting}

The file is structured by certain keywords which we explain in the following:
\begin{description}
\item{\verb$version$:} The version of \texttt{AutoEFT} which
  was used to produce the output file.

\item {\verb$type$:} A list\footnote{In \texttt{YAML}, this is called a
\textit{sequence}. Its elements are marked by the leading dashes in lines~4 and 5 of
\cref{lst:dim10}.} of two elements. The first element (line~4) specifies the
  multiplicities of the fields and derivatives in the operator. It is
  equivalent to the specification in \cref{eq:kush}.  The second element
  (line~5) states that the operator type is \verb$complex$, (i.e.~the
  operators are not Hermitian). The Hermitian conjugate type is contained in a
  different file. In the present case, this would be
  \verb$1WL_1H+_1Q+_1uC+_1WL+_2D.yml$. For Hermitian operators, the second
  entry of \verb$type$ is \verb$real$.

\item{\verb$generations$:} Provides the number of generations for each
  field. In this case, it specifies that there is only a single generation of
  \verb$WL$, \verb$WL+$, and \verb$H$, but three generations of quarks
  \verb$Q$ and \verb$uC$.

\item{\verb$n_terms$:} The total number of operators with independent Lorentz
  and $\SU(n)$ index contractions.  It does not take into account the number
  of generations though.  In \cref{lst:dim10}, there are five independent ways
  of contracting the Lorentz indices of the operators while there is only one
  option for $\SU(3)$, and two for $\SU(2)$, so the number of terms \verb$n_terms$
  is $5\cdot 1\cdot 2=10$.

\item{\verb$n_operators$:} The total number of independent operators, taking
  into account the independent values the generation indices can take.  For
  \cref{eq:kush}, there are $3 \cdot 3 = 9$ independent combinations of the
  $Q$ and $u_\mathbb{C}$ generation indices for each of the ten
  terms, hence \verb$n_operators$ is $10 \cdot 9 = 90$.

\item{\verb$invariants$:} The list of invariant index contractions of the
  fields in the operator. The contraction of indices for each internal
  symmetry group and for the Lorentz group is listed separately, as indicated
  by the sub-keywords \verb$Lorentz$, \verb$SU(3)$, and \verb$SU(2)$ in
  lines~10, 16, and 18. Each independent contraction is labeled by
  \verb$O(<G>,<m>)$, where \verb$<G>$ is the name of the Lorentz or internal
  symmetry group, and \verb$<m>$ enumerates the contractions.
  
  The indices are denoted by \verb$<i>_<j>$, where \verb$<i>$
  is the position of the field that carries this index, and
  \verb$<j>$ is the position of the index \textit{on} the field.  Per
  invariant contraction, each index appears exactly twice, and summation is
  implied.  For the Lorentz group, \verb$<i>_<j>~$
  denotes dotted indices. Furthermore, the indices are associated with the
  building blocks of \cref{eq:edom}, rather than with the fields.  Note that
  the dotted and undotted indices of the building blocks are understood to be
  (separately) symmetrized (cf.~\cref{sec:theory:redundancies}).

  The symbol \verb$eps$ denotes the $\epsilon$-tensor with
  \verb$eps(1,2)=eps(2~,1~)=1$ for the Lorentz
  group, and \verb$eps(1,2,...,n)=1$ for any internal $\SU(n)$ group. All
  indices not associated with the symmetry group in question are suppressed
  on the fields.\footnote{This means that, if the operators contain only fields that are singlets under a
  particular symmetry group \texttt{G} (i.e.~if there is no index to be
  contracted), the corresponding entry \texttt{G:} contains just a single
  element \texttt{+1} multiplied by the fields without indices.}

\item{\verb$permutation_symmetries$:} The list of permutation symmetries
  according to \cref{sec:theory:permutation}. Since the operator type of
  \cref{eq:kush} does not involve repeated fields, this entry (lines~22--36)
  is redundant and is included only for consistency. Details will be discussed
  in \cref{sec:repeat} for a non-trivial example.
\end{description}

In \cref{lst:dim10}, the invariant contractions are given by
(cf.~\cref{eq:lorentz,eq:gauge})
\begin{equation}\label{eq:glow}
  \begin{aligned}
    &\text{line 11:}\quad
    \mathcal{O}^{\text{Lorentz}}_{1} =
    T_1^\text{Lorentz}\circ \mathcal{Q}_\text{NRF}=
    \\&\qquad=
\epsilon^{\alpha_1\gamma_1}\,\epsilon^{\alpha_2\gamma_2}\,\epsilon^{\beta_1\gamma_3}\,\epsilon_{\dot{\gamma}_1\dot{\eta}_1}\,\epsilon_{\dot{\gamma}_2\dot{\eta}_2}
\, {W_\mathrm{L}}_{\alpha_1\alpha_2}\,Q_{\beta_1}\,(D^2
u_\mathbb{C})_{(\gamma_1\gamma_2\gamma_3)}^{(\dot{\gamma}_1\dot{\gamma}_2)}\,
H\,{W^\dagger_\mathrm{L}}^{\dot{\eta}_1\dot{\eta}_2}\,,
    \\[.5em]
    &\text{line 12:}\quad
    \mathcal{O}^{\text{Lorentz}}_{2} =
    T_2^\text{Lorentz}\circ \mathcal{Q}_\text{NRF}=
    \\&\qquad=-\epsilon^{\alpha_1\gamma_1}\,\epsilon^{\alpha_2\gamma_2}\,\epsilon^{\beta_1\delta_1}\,\epsilon_{\dot{\gamma}_1\dot{\eta}_1}\,\epsilon_{\dot{\delta}_1\dot{\eta}_2}
    \, {W_\mathrm{L}}_{\alpha_1\alpha_2}\,Q_{\beta_1}\,(D
    u_\mathbb{C})_{(\gamma_1\gamma_2)}^{\dot{\gamma}_1}
    \,(D H)_{\delta_1}^{\dot{\delta}_1}\,{W^\dagger_\mathrm{L}}^{\dot{\eta}_1\dot{\eta}_2}\,,
    \\[.5em]
    &\text{line 13:}\quad
        \mathcal{O}^{\text{Lorentz}}_{3} =
    T_3^\text{Lorentz}\circ \mathcal{Q}_\text{NRF}=
    \\&\qquad=
    \epsilon^{\alpha_1\gamma_1}\,\epsilon^{\alpha_2\delta_1}\,\epsilon^{\beta_1\delta_2}\,\epsilon_{\dot{\delta}_1\dot{\eta}_1}\,\epsilon_{\dot{\delta}_2\dot{\eta}_2} \, {W_\mathrm{L}}_{\alpha_1\alpha_2}\,Q_{\beta_1}\,{u_\mathbb{C}}_{\gamma_1}\,(D^2 H)_{(\delta_1\delta_2)}^{(\dot{\delta}_1\dot{\delta}_2)}\,{W^\dagger_\mathrm{L}}^{\dot{\eta}_1\dot{\eta}_2}\,,
    \\[.5em]
    &\text{line 14:}\quad
        \mathcal{O}^{\text{Lorentz}}_{4} =
    T_4^\text{Lorentz}\circ \mathcal{Q}_\text{NRF}=
    \\&\qquad=
    -\epsilon^{\alpha_1\beta_1}\,\epsilon^{\alpha_2\gamma_1}\,\epsilon^{\gamma_2\delta_1}\,\epsilon_{\dot{\gamma}_1\dot{\eta}_1}\,\epsilon_{\dot{\delta}_1\dot{\eta}_2}
    \, {W_\mathrm{L}}_{\alpha_1\alpha_2}\,Q_{\beta_1}\,(D
    u_\mathbb{C})_{(\gamma_1\gamma_2)}^{\dot{\gamma}_1}
    \,(D H)_{\delta_1}^{\dot{\delta}_1}\,{W^\dagger_\mathrm{L}}^{\dot{\eta}_1\dot{\eta}_2}\,,
    \\[.5em]
    &\text{line 15:}\quad
        \mathcal{O}^{\text{Lorentz}}_{5} =
    T_5^\text{Lorentz}\circ \mathcal{Q}_\text{NRF}=
    \\&\qquad=
    \epsilon^{\alpha_1\beta_1}\,\epsilon^{\alpha_2\delta_1}\,\epsilon^{\gamma_1\delta_2}\,\epsilon_{\dot{\delta}_1\dot{\eta}_1}\,\epsilon_{\dot{\delta}_2\dot{\eta}_2} \, {W_\mathrm{L}}_{\alpha_1\alpha_2}\,Q_{\beta_1}\,{u_\mathbb{C}}_{\gamma_1}\,(D^2 H)_{(\delta_1\delta_2)}^{(\dot{\delta}_1\dot{\delta}_2)}\,{W^\dagger_\mathrm{L}}^{\dot{\eta}_1\dot{\eta}_2}\,,
    \\[1em]
    &\text{line 17:}\quad\mathcal{O}^{\SU(3)}_{1} =
T^{\SU(3)}_1\circ\mathcal{Q}_\text{NRF}=
\epsilon^{b_1c_1c_2} \,
W_\mathrm{L}\,Q_{b_1}\,{u_\mathbb{C}}_{c_1c_2}\,H\,W_\mathrm{L}^\dagger\,,
\\[1em]
&\text{line 19:}\quad\mathcal{O}^{\SU(2)}_{1} =
T^{\SU(2)}_1\circ\mathcal{Q}_\text{NRF}=
\epsilon^{k_1m_1}\,\epsilon^{k_2n_1}\,\epsilon^{l_1n_2} \,
{W_\mathrm{L}}_{k_1k_2}\,Q_{l_1}\,u_\mathbb{C}\,H_{m_1}\,{W^\dagger_\mathrm{L}}_{n_1n_2}\,,
\\[.5em]
&\text{line 20:}\quad\mathcal{O}^{\SU(2)}_{2} =
T^{\SU(2)}_2\circ\mathcal{Q}_\text{NRF}=
\epsilon^{k_1l_1}\,\epsilon^{k_2n_1}\,\epsilon^{m_1n_2}
\, {W_\mathrm{L}}_{k_1k_2}\,Q_{l_1}\,u_\mathbb{C}\,H_{m_1}\,{W^\dagger_\mathrm{L}}_{n_1n_2}\,,
  \end{aligned}
\end{equation}
with $\mathcal{Q}_\text{NRF}$ from \cref{eq:kush}. The operation
$T\circ\mathcal{Q}$ implies that all indices other than those of $T$ are
suppressed in $\mathcal{Q}$. Since only contractions with the Lorentz tensors
determine the positions of the derivatives, we omit the latter in the
contractions with tensors of the internal symmetries.

The complete set of independent operators $\boldsymbol{T}\cdot\mathcal{Q}_\text{NRF}$
is thus given by all possible combinations of the Lorentz and $\SU(n)$ tensors:
\begin{equation}\label{eq:OOO}
  \boldsymbol{T}\equiv T^{\text{Lorentz}} \otimes T^{\SU(3)} \otimes T^{\SU(2)}
    =
    \begin{psmallmatrix}
         T^{\text{Lorentz}}_1\,\otimes\, T^{\SU(3)}_1\,\otimes\, T^{\SU(2)}_1 \\
         T^{\text{Lorentz}}_1\,\otimes\, T^{\SU(3)}_1\,\otimes\, T^{\SU(2)}_2 \\
         T^{\text{Lorentz}}_2\,\otimes\, T^{\SU(3)}_1\,\otimes\, T^{\SU(2)}_1 \\
         T^{\text{Lorentz}}_2\,\otimes\, T^{\SU(3)}_1\,\otimes\, T^{\SU(2)}_2 \\
         T^{\text{Lorentz}}_3\,\otimes\, T^{\SU(3)}_1\,\otimes\, T^{\SU(2)}_1 \\
         T^{\text{Lorentz}}_3\,\otimes\, T^{\SU(3)}_1\,\otimes\, T^{\SU(2)}_2 \\
         T^{\text{Lorentz}}_4\,\otimes\, T^{\SU(3)}_1\,\otimes\, T^{\SU(2)}_1 \\
         T^{\text{Lorentz}}_4\,\otimes\, T^{\SU(3)}_1\,\otimes\, T^{\SU(2)}_2 \\
         T^{\text{Lorentz}}_5\,\otimes\, T^{\SU(3)}_1\,\otimes\, T^{\SU(2)}_1 \\
         T^{\text{Lorentz}}_5\,\otimes\, T^{\SU(3)}_1\,\otimes\, T^{\SU(2)}_2 \\
    \end{psmallmatrix}
    \,.
\end{equation}
Three examples for the ten independent terms (cf.\,line~7 in \cref{lst:dim10}) are given by
\begin{equation}\label{eq:flew}
  \begin{aligned}
    (T^{\text{Lorentz}}_{1} \otimes
    T^{\SU(3)}_{1}& \otimes T^{\SU(2)}_{1})\circ\mathcal{Q}_\text{NRF}=\\
    &=
    \epsilon^{\alpha_1\gamma_1}\,\epsilon^{\alpha_2\gamma_2}\,\epsilon^{\beta_1\gamma_3}\,\epsilon_{\dot{\gamma}_1\dot{\eta}_1}\,\epsilon_{\dot{\gamma}_2\dot{\eta}_2}\,
\epsilon^{b_1c_1c_2} \,
\epsilon^{k_1m_1}\,\epsilon^{k_2n_1}\,\epsilon^{l_1n_2} \,\\
&\qquad\times
    {W_\mathrm{L}}_{\alpha_1\alpha_2k_1k_2}\,Q_{\beta_1b_1l_1}\,(D^2
    {u_\mathbb{C}}_{c_1c_2})_{(\gamma_1\gamma_2\gamma_3)}^{(\dot{\gamma}_1\dot{\gamma}_2)}\,
H_{m_1}\,{W^\dagger_\mathrm{L}}_{n_1n_2}^{\dot{\eta}_1\dot{\eta}_2}\,,\\[.5em]
    (T^{\text{Lorentz}}_{1} \otimes
    T^{\SU(3)}_{1}& \otimes T^{\SU(2)}_{2})\circ\mathcal{Q}_\text{NRF}=\\
    &=
    \epsilon^{\alpha_1\gamma_1}\,\epsilon^{\alpha_2\gamma_2}\,\epsilon^{\beta_1\gamma_3}\,\epsilon_{\dot{\gamma}_1\dot{\eta}_1}\,\epsilon_{\dot{\gamma}_2\dot{\eta}_2}\,
\epsilon^{b_1c_1c_2} \,
\epsilon^{k_1l_1}\,\epsilon^{k_2n_1}\,\epsilon^{m_1n_2}
\,\\
&\qquad\times
    {W_\mathrm{L}}_{\alpha_1\alpha_2k_1k_2}\,Q_{\beta_1b_1l_1}\,(D^2
    {u_\mathbb{C}}_{c_1c_2})_{(\gamma_1\gamma_2\gamma_3)}^{(\dot{\gamma}_1\dot{\gamma}_2)}\,
H_{m_1}\,{W^\dagger_\mathrm{L}}_{n_1n_2}^{\dot{\eta}_1\dot{\eta}_2}\,,
\\[.5em]
(T^{\text{Lorentz}}_{2} \otimes
    T^{\SU(3)}_{1}& \otimes T^{\SU(2)}_{1})\circ\mathcal{Q}_\text{NRF}=\\
    &=
    -\epsilon^{\alpha_1\gamma_1}\,\epsilon^{\alpha_2\gamma_2}\,\epsilon^{\beta_1\delta_1}\,\epsilon_{\dot{\gamma}_1\dot{\eta}_1}\,\epsilon_{\dot{\delta}_1\dot{\eta}_2}\,
\epsilon^{b_1c_1c_2} \,
\epsilon^{k_1m_1}\,\epsilon^{k_2n_1}\,\epsilon^{l_1n_2} \,\\
&\qquad\times
    \, {W_\mathrm{L}}_{\alpha_1\alpha_2k_1k_2}\,Q_{\beta_1b_1l_1}\,(D
    {u_\mathbb{C}}_{c_1c_2})_{(\gamma_1\gamma_2)}^{\dot{\gamma}_1}
    \,(D H_{m_1})_{\delta_1}^{\dot{\delta}_1}\,{W^\dagger_\mathrm{L}}_{n_1n_2}^{\dot{\eta}_1\dot{\eta}_2}\,.
  \end{aligned}
\end{equation}
Up to this point, we have suppressed the generation indices of the
quarks. Since quark and anti-quark transform differently under the symmetry
groups, we can simply attach a generation index to each of the two quark fields
in \cref{eq:kush}. 


\subsection{Repeated Fields}\label{sec:repeat}

The operator \cref{eq:kush} treated in \cref{sec:norepeat} is special in the
sense that it does not contain repeated fields, i.e.~all fields in this type
transform differently under the symmetry group of the theory.  As mentioned in
\cref{sec:theory:permutation}, if repeated fields are present, new redundancies
can arise. The output file thus requires information on the permutation
symmetry. As an example, we consider the dimension-11 operators of type
\begin{equation}\label{eq:norepeat:iago}
  \begin{aligned}
    \mathcal{Q}_\text{RF} &= L^2 d_\mathbb{C} e_\mathbb{C} u^2_\mathbb{C} H^2\,,
  \end{aligned}
\end{equation}
contained in the file
\begin{verbatim}
  operators/8/6psi_2phi/2L_1dC_1eC_2uC_2H.yml
\end{verbatim}
that is included in \verb$smeft_11.tar.xz$. Its content is displayed
in \cref{lst:dim11}.
\begin{lstlisting}[caption={Content of file
    \texttt{2L\symbol{95}1dC\symbol{95}1eC\symbol{95}2uC\symbol{95}2H.yml},
    which encodes the operators of \cref{eq:norepeat:iago}.
  },
  label={lst:dim11}, style=yaml]
# '2L_1dC_1eC_2uC_2H.yml' generated by AutoEFT 1.0.0
version: 1.0.0
type:
- {L: 2, dC: 1, eC: 1, uC: 2, H: 2}
- complex
generations: {L: 3, dC: 3, eC: 3, uC: 3, H: 1}
n_terms: 5
n_operators: 891
invariants:
  Lorentz:
    O(Lorentz,1): +eps(1_1,4_1)*eps(2_1,5_1)*eps(3_1,6_1) * L(1_1)*L(2_1)*dC(3_1)*eC(4_1)*uC(5_1)*uC(6_1)*H*H
    O(Lorentz,2): +eps(1_1,3_1)*eps(2_1,5_1)*eps(4_1,6_1) * L(1_1)*L(2_1)*dC(3_1)*eC(4_1)*uC(5_1)*uC(6_1)*H*H
    O(Lorentz,3): +eps(1_1,3_1)*eps(2_1,4_1)*eps(5_1,6_1) * L(1_1)*L(2_1)*dC(3_1)*eC(4_1)*uC(5_1)*uC(6_1)*H*H
    O(Lorentz,4): +eps(1_1,2_1)*eps(3_1,5_1)*eps(4_1,6_1) * L(1_1)*L(2_1)*dC(3_1)*eC(4_1)*uC(5_1)*uC(6_1)*H*H
    O(Lorentz,5): +eps(1_1,2_1)*eps(3_1,4_1)*eps(5_1,6_1) * L(1_1)*L(2_1)*dC(3_1)*eC(4_1)*uC(5_1)*uC(6_1)*H*H
  SU3:
    O(SU3,1): +eps(3_1,3_2,5_2)*eps(5_1,6_1,6_2) * L*L*dC(3_1,3_2)*eC*uC(5_1,5_2)*uC(6_1,6_2)*H*H
  SU2:
    O(SU2,1): +eps(1_1,7_1)*eps(2_1,8_1) * L(1_1)*L(2_1)*dC*eC*uC*uC*H(7_1)*H(8_1)
    O(SU2,2): +eps(1_1,2_1)*eps(7_1,8_1) * L(1_1)*L(2_1)*dC*eC*uC*uC*H(7_1)*H(8_1)
permutation_symmetries:
- vector: Lorentz * SU3 * SU2
- symmetry: {L: [1, 1], dC: [1], eC: [1], uC: [2], H: [2]}
  n_terms: 2
  n_operators: 324
  matrix: |-
    [ 8 -4  0  0 -4  2 -4  2  6 -3]
    [ 0  0  8 -4 -4  2 -4  2  2 -1]
- symmetry: {L: [1, 1], dC: [1], eC: [1], uC: [1, 1], H: [2]}
  n_terms: 1
  n_operators: 81
  matrix: |-
    [ 0  0  0  0  4 -2  0  0 -2  1]
- symmetry: {L: [2], dC: [1], eC: [1], uC: [2], H: [2]}
  n_terms: 1
  n_operators: 324
  matrix: |-
    [ 0  0  0  0  0  0  4 -2 -2  1]
- symmetry: {L: [2], dC: [1], eC: [1], uC: [1, 1], H: [2]}
  n_terms: 1
  n_operators: 162
  matrix: |-
    [ 0  0  0  0  0  0  0  0 -2  1]
\end{lstlisting}

The meaning of the first 20 lines in this file was explained in the previous
section.  If all fields in \cref{eq:norepeat:iago} were different, one would
arrive at 10 terms, according to the $5\cdot 1\cdot 2 = 10$ independent Lorentz and
$\SU(n)$ tensors:
\begin{equation}\label{eq:OO}
  \boldsymbol{T}\equiv T^{\text{Lorentz}} \otimes T^{\SU(3)} \otimes T^{\SU(2)}
    =
    \begin{psmallmatrix}
         T^{\text{Lorentz}}_1\,\otimes\, T^{\SU(3)}_1\,\otimes\, T^{\SU(2)}_1 \\
         T^{\text{Lorentz}}_1\,\otimes\, T^{\SU(3)}_1\,\otimes\, T^{\SU(2)}_2 \\
         T^{\text{Lorentz}}_2\,\otimes\, T^{\SU(3)}_1\,\otimes\, T^{\SU(2)}_1 \\
         T^{\text{Lorentz}}_2\,\otimes\, T^{\SU(3)}_1\,\otimes\, T^{\SU(2)}_2 \\
         T^{\text{Lorentz}}_3\,\otimes\, T^{\SU(3)}_1\,\otimes\, T^{\SU(2)}_1 \\
         T^{\text{Lorentz}}_3\,\otimes\, T^{\SU(3)}_1\,\otimes\, T^{\SU(2)}_2 \\
         T^{\text{Lorentz}}_4\,\otimes\, T^{\SU(3)}_1\,\otimes\, T^{\SU(2)}_1 \\
         T^{\text{Lorentz}}_4\,\otimes\, T^{\SU(3)}_1\,\otimes\, T^{\SU(2)}_2 \\
         T^{\text{Lorentz}}_5\,\otimes\, T^{\SU(3)}_1\,\otimes\, T^{\SU(2)}_1 \\
         T^{\text{Lorentz}}_5\,\otimes\, T^{\SU(3)}_1\,\otimes\, T^{\SU(2)}_2 \\
    \end{psmallmatrix}
    \,.
\end{equation}
Attaching generation indices to the fields, the number of operators would be $10\cdot
3^6=7290$, because there are six fields with three generations each.
This, however, neglects the permutation symmetries arising from the repeated fields.
In this example, there are three sets of repeated fields, each of which occurs twice, as can be
seen from the first element of \verb$type$ (line~4).  This reduces the actual
number of terms and operators to those given in lines 7 and 8.
As described in \cref{sec:theory:permutation}, this is because, on the one
hand, the permutation symmetries induce dependencies between the terms in \cref{eq:OO},
reducing the number of independent terms. On the other hand,
a particular permutation symmetry restricts the number of independent
combinations of generation indices in the field monomials.

Consider, for example, the
second item (marked by the character ``\verb$-$'' in the first column) under
\verb$permutation_symmetries$, i.e.~lines~23--28 of \cref{lst:dim11}.
The entry for the keyword \verb$symmetry$ should be
interpreted as a set of Young tableaux, representing the permutation symmetry
for each kind of field. In this
case\footnote{I.e.~\texttt{[i,j,\ldots]}  in the \texttt{symmetry} entry denotes
a Young diagram with \texttt{i} boxes in the first row, \texttt{j} boxes in
the second, etc.}
\begin{equation}\label[relation]{eq:norepeat:judd}
  \begin{aligned}
    &\mbox{\texttt{L:}}\mbox{\texttt{[1,1]}}&\quad\hat{=}\quad
    &&\lambda_L &= \begin{ytableau}i\\j\end{ytableau}\,,\\[.2em]
    &\mbox{\texttt{dC:}}\mbox{\texttt{ [1]}}&\quad\hat{=}\quad
    &&\lambda_{d_\mathbb{C}} &= \begin{ytableau}k\end{ytableau}\,,\\[.5em]
    &\mbox{\texttt{eC:}}\mbox{\texttt{ [1]}}&\quad\hat{=}\quad
    &&\lambda_{e_\mathbb{C}} &= \begin{ytableau}l\end{ytableau}\,,\\[.5em]
    &\mbox{\texttt{uC:}}\mbox{\texttt{ [2]}}&\quad\hat{=}\quad
    &&\lambda_{u_\mathbb{C}} &= \begin{ytableau}m&n\end{ytableau}\,,\\[.5em]
    &\mbox{\texttt{H:}}\mbox{\texttt{ [2]}}&\quad\hat{=}\quad
    &&\lambda_H &= \begin{ytableau}r&s\end{ytableau}\,.
  \end{aligned}
\end{equation}
The fillings are the generation indices, i.e.~$i,j,k,l,m,n\in \set{1,2,3}$, and
$r=s=1$, because there are three generations of leptons and quarks, and only
one of Higgs bosons.

For this permutation symmetry, only the contractions with two combinations
of the ten tensors in \cref{eq:OO} are independent. They are given by
(cf.\ \cref{eq:permutation})
\begin{equation}\label{eq:smeft:chaw}
  \begin{aligned}
    \boldsymbol{\mathcal{T}} = \mathcal{K}\boldsymbol{T}\,,
  \end{aligned}
\end{equation}
where
\begin{equation}\label{eq:norepeat:jhwh}
  \begin{aligned}
                \mathcal{K} =
            \begin{pmatrix}
                8 & -4 & 0 & 0 & -4 & 2 & -4 & 2 & 6 & -3 \\
                0 & 0 & 8 & -4 & -4 & 2 & -4 & 2 & 2 & -1
            \end{pmatrix}
  \end{aligned}
\end{equation}
is the matrix listed under the keyword \verb$matrix$ in lines~26--28 of
\cref{lst:dim11}. The rank of this matrix is equal to the number of terms
\verb$n_terms$ in line~24. Its form depends on the order of the factors in the
Kronecker product\footnote{The Kronecker product $A\otimes B$ of an $m\times
n$ matrix $A$ with a $p\times q$ matrix $B$ is a $pm\times qn$ matrix which is
obtained by replacing all entries of $A$ by their product with the matrix
$B$. Note that $A\otimes (B\otimes C)=(A\otimes B)\otimes C\equiv A\otimes
B\otimes C$.} in \cref{eq:OO}. For clarity, this order is added to the
\texttt{AutoEFT} output file under the keyword \verb$vector$, see line~22 in
\cref{lst:dim11}.

Considering the permutation symmetries of the fields, only such values of generation indices
are independent for which the Young tableaux in \cref{eq:norepeat:judd} are
semi-standard. For the case at hand, we thus have
\begin{equation}\label{eq::golp}
  \begin{aligned}
    (i,j)&\in\set{(1,2),(1,3),(2,3)} \ \equiv\ \mathcal{S}^{(3)}_{1,1}\,,\\
    (m,n)&\in\set{(1,1),(1,2),(1,3),(2,2),(2,3),(3,3)}
    \ \equiv\ \mathcal{S}^{(3)}_{2}\,,\\
    (r,s)&\in\set{(1,1)} \ \equiv\ \mathcal{S}^{(1)}_{2}\,,
  \end{aligned}
\end{equation}
where $\mathcal{S}_{\lambda_\Phi}^{(n_g)}$ is the set of independent
generation indices for the repeated field $\Phi$ with permutation symmetry
$\lambda_\Phi$ and $n_g$ generations.  The generation indices of fields that
appear only once are still independent and are therefore not restricted
further, i.e.~$k,l\in\set{1,2,3}$.  In total, out of the $3^6$ combinations of
the generation indices, only $3\cdot 6\cdot 1\cdot 3^2 = 162$ need to be
considered for this particular representation of the permutation symmetry. 
There are thus $2\cdot(3\cdot 6\cdot 1\cdot 3^2) = 324$ independent operators
of this type, in agreement with line~25.

There are three more permutation symmetries:
\begin{itemize}
\item lines~29--33:
  $\lambda_L = \ytableaushort{i,j}\,,\
  \lambda_{d_\mathbb{C}} = \ytableaushort{k}\,,\
  \lambda_{e_\mathbb{C}} = \ytableaushort{l}\,,\
  \lambda_{u_{\mathbb{C}}} = \ytableaushort{m,n}\,,\
  \lambda_{H} = \ytableaushort{rs}$,\\
  in which case
  there is only one independent combination (cf.~line~30) of tensor
  contractions, with
  \begin{equation}\label{eq:norepeat:inky}
    \begin{aligned}
            \mathcal{K} &=
            \begin{pmatrix}
                0 & 0 & 0 & 0 & 4 & -2 & 0 & 0 & -2 & 1
            \end{pmatrix}\,.
    \end{aligned}
  \end{equation}
  The independent values of generation indices are given by $(i,j),(m,n)\in\mathcal{S}_{1,1}^{(3)}$,
  $(r,s)\in\mathcal{S}_{2}^{(1)}$,
  and $k,l\in\set{1,2,3}$.   This results in $1\cdot(3\cdot 3\cdot1\cdot 3^2)=81$ independent operators, as
  indicated in line~31.

\item lines~34--38:
  $\lambda_L = \ytableaushort{ij}\,,\
  \lambda_{d_\mathbb{C}} = \ytableaushort{k}\,,\
  \lambda_{e_\mathbb{C}} = \ytableaushort{l}\,,\
  \lambda_{u_{\mathbb{C}}} = \ytableaushort{mn}\,,\
  \lambda_{H} = \ytableaushort{rs}$,\\
  in which case
  the only independent term is given by the matrix
    \begin{equation}\label{eq:norepeat:inky2}
        \mathcal{K} =
        \begin{pmatrix}
            0 & 0 & 0 & 0 & 0 & 0 & 4 & -2 & -2 & 1
        \end{pmatrix}\,,
    \end{equation}
  and with $(i,j),(m,n)\in\mathcal{S}_{2}^{(3)}$,
  $(r,s)\in\mathcal{S}_{2}^{(1)}$,
  and $k,l\in\set{1,2,3}$, resulting in $1\cdot(6\cdot6\cdot1\cdot3^2)=324$ operators.
\item lines~39--43:
  $\lambda_L = \ytableaushort{ij}\,,\
  \lambda_{d_\mathbb{C}} = \ytableaushort{k}\,,\
  \lambda_{e_\mathbb{C}} = \ytableaushort{l}\,,\
  \lambda_{u_{\mathbb{C}}} = \ytableaushort{m,n}\,,\
  \lambda_{H} = \ytableaushort{rs}$,\\
  in which case
  the only independent term is given by the matrix
    \begin{equation}\label{eq:norepeat:ink3}
        \mathcal{K} =
        \begin{pmatrix}
            0 & 0 & 0 & 0 & 0 & 0 & 0 & 0 & -2 & 1
        \end{pmatrix}\,,
    \end{equation}
  and with $(i,j)\in\mathcal{S}_{2}^{(3)}$,
  $(m,n)\in\mathcal{S}_{1,1}^{(3)}$,
  $(r,s)\in\mathcal{S}_{2}^{(1)}$,
  and $k,l\in\set{1,2,3}$, resulting in $1\cdot(6\cdot 3\cdot1\cdot3^2)=162$ operators.
\end{itemize}
In total, the file \verb$2L_1dC_1eC_2uC_2H.yml$ thus encodes
$324+81+324+162=891$ operators, as stated in line~8 of
\cref{lst:dim11}. The number of terms \verb$n_terms$ in line~7 indicates that there
are five independent combinations, given by
\cref{eq:norepeat:jhwh,eq:norepeat:inky,eq:norepeat:inky2,eq:norepeat:ink3}.


\subsection{SMEFT at Mass Dimension 12}\label{sec:dim12}

The complete basis of operators up to mass dimension~12 for \smeft\ is
provided, in the format described in \cref{sec:norepeat,sec:repeat}, in the
ancillary files together with this paper. It took a few seconds to generate
the operator basis at mass dimensions 5, 6, and 7, a few minutes for mass
dimension~8 and~9, and a few hours for mass dimension~10 and~11. At mass
dimension~12, the largest amount of time was spent by \texttt{AutoEFT} on
incorporating the permutation symmetry of the operators with six gluonic field
strength tensors, which takes of the order of $10^4$ \abbrev{CPU} hours.  For
example, for the type $G_{\mathrm{L}}^6$, the $2175$ contractions of the field
strength tensors with the 15 independent Lorentz and 145 independent $\SU(3)$
tensors is reduced to only eight independent operators. Their explicit
expressions are too long to display them in this paper. They are encoded in
the file
\begin{verbatim}
  operators/6/6FL/6GL.yml
\end{verbatim}
which is about 140\,\abbrev{KB} in size.  Instead, as an example of a
dimension-12 operator, we display the
following type of eight-fermion operators in \cref{lst:smeft:ahem}:
\begin{equation}\label[type]{eq:smeft:ahem}
  \begin{aligned}
      L^3 e_\mathbb{C} L^\dagger d_\mathbb{C}^{\dagger\,2} u_\mathbb{C}^\dagger \,,
  \end{aligned}
\end{equation}
contained in the file
\begin{verbatim}
  operators/8/4psi_4psi+/3L_1eC_1L+_2dC+_1uC+.yml
\end{verbatim}
that is included in \verb$smeft_12.tar.xz$.
\begin{lstlisting}[caption={Content of file
    \texttt{3L\symbol{95}1eC\symbol{95}1L+\symbol{95}2dC+\symbol{95}1uC+.yml},
    which encodes the operators of \cref{eq:smeft:ahem}.
  },
  label={lst:smeft:ahem}, style=yaml]
# '3L_1eC_1L+_2dC+_1uC+.yml' generated by AutoEFT 1.0.0
version: 1.0.0
type:
- {L: 3, eC: 1, L+: 1, dC+: 2, uC+: 1}
- complex
generations: {L: 3, eC: 3, L+: 3, dC+: 3, uC+: 3}
n_terms: 6
n_operators: 4617
invariants:
  Lorentz:
    O(Lorentz,1): +eps(1_1,3_1)*eps(2_1,4_1)*eps(5_1~,7_1~)*eps(6_1~,8_1~) * L(1_1)*L(2_1)*L(3_1)*eC(4_1)*L+(5_1~)*dC+(6_1~)*dC+(7_1~)*uC+(8_1~)
    O(Lorentz,2): +eps(1_1,3_1)*eps(2_1,4_1)*eps(5_1~,6_1~)*eps(7_1~,8_1~) * L(1_1)*L(2_1)*L(3_1)*eC(4_1)*L+(5_1~)*dC+(6_1~)*dC+(7_1~)*uC+(8_1~)
    O(Lorentz,3): +eps(1_1,2_1)*eps(3_1,4_1)*eps(5_1~,7_1~)*eps(6_1~,8_1~) * L(1_1)*L(2_1)*L(3_1)*eC(4_1)*L+(5_1~)*dC+(6_1~)*dC+(7_1~)*uC+(8_1~)
    O(Lorentz,4): +eps(1_1,2_1)*eps(3_1,4_1)*eps(5_1~,6_1~)*eps(7_1~,8_1~) * L(1_1)*L(2_1)*L(3_1)*eC(4_1)*L+(5_1~)*dC+(6_1~)*dC+(7_1~)*uC+(8_1~)
  SU3:
    O(SU3,1): +eps(6_1,7_1,8_1) * L*L*L*eC*L+*dC+(6_1)*dC+(7_1)*uC+(8_1)
  SU2:
    O(SU2,1): +eps(1_1,3_1)*eps(2_1,5_1) * L(1_1)*L(2_1)*L(3_1)*eC*L+(5_1)*dC+*dC+*uC+
    O(SU2,2): +eps(1_1,2_1)*eps(3_1,5_1) * L(1_1)*L(2_1)*L(3_1)*eC*L+(5_1)*dC+*dC+*uC+
permutation_symmetries:
- vector: Lorentz * SU3 * SU2
- symmetry: {L: [1, 1, 1], eC: [1], L+: [1], dC+: [2], uC+: [1]}
  n_terms: 1
  n_operators: 162
  matrix: |-
    [ 2 -1  2 -1 -1  2 -1  2]
- symmetry: {L: [2, 1], eC: [1], L+: [1], dC+: [2], uC+: [1]}
  n_terms: 1
  n_operators: 1296
  matrix: |-
    [ 0  1  0  1  1 -1  1 -1]
- symmetry: {L: [3], eC: [1], L+: [1], dC+: [2], uC+: [1]}
  n_terms: 1
  n_operators: 1620
  matrix: |-
    [ 0  1  0  1 -1  0 -1  0]
- symmetry: {L: [1, 1, 1], eC: [1], L+: [1], dC+: [1, 1], uC+: [1]}
  n_terms: 1
  n_operators: 81
  matrix: |-
    [ 2 -1 -2  1 -1  2  1 -2]
- symmetry: {L: [2, 1], eC: [1], L+: [1], dC+: [1, 1], uC+: [1]}
  n_terms: 1
  n_operators: 648
  matrix: |-
    [ 0  1  0 -1  1 -1 -1  1]
- symmetry: {L: [3], eC: [1], L+: [1], dC+: [1, 1], uC+: [1]}
  n_terms: 1
  n_operators: 810
  matrix: |-
    [ 0  1  0 -1 -1  0  1  0]
\end{lstlisting}

In this example, there are two sets of repeated fields, $L^3$ and
$d_\mathbb{C}^{\dagger\,2}$.  Lines~22, 27, 32, 37, 42, and~47 specify the
permutation symmetries of the generation indices $i,j,k$ of the three lepton
doublets
\begin{equation}\label{eq:sym:L3}
    \lambda_L \in
    \Set{
    \ytableaushort{i,j,k}
    \,,\,
    \ytableaushort{ij,k}
    \,,\,
    \ytableaushort{ijk}
    }\,.
\end{equation}
The corresponding sets of independent generation indices are thus given by
\begin{equation}
    \begin{aligned}
        \mathcal{S}_{1,1,1}^{(3)} &\equiv \set{(1,2,3)} \,,\\
        \mathcal{S}_{2,1}^{(3)} &\equiv \set{(i,j,k)|1\leq i\leq j\leq
          3\ \wedge\ i<k\leq 3}\,,\\
        \mathcal{S}_{3}^{(3)} &\equiv \set{(i,j,k)|1\leq i\leq j\leq k\leq
          3}\,,
    \end{aligned}
\end{equation}
containing 1, 8, and 10 elements, respectively.  The permutation symmetries of
the generation indices $m,n$ of the two down-type quarks are given by
\begin{equation}\label{eq:sym:d2}
    \lambda_{d_\mathbb{C}} \in
    \Set{
    \ytableaushort{mn}
    \,,\,
    \ytableaushort{m,n}
    }\,,
\end{equation}
and thus the corresponding sets of generation indices are
$\mathcal{S}_{2}^{(3)}$ and $\mathcal{S}_{1,1}^{(3)}$ with 6 and 3 elements,
respectively (see \cref{eq::golp}).

In \cref{lst:smeft:ahem}, all combinations of the symmetries
\cref{eq:sym:L3,eq:sym:d2} are present, and each combination is given by
exactly one term.  Therefore, including the $3^3$ generation multiplicities of $e_\mathbb{C}$, $L^\dagger$,
and $u_\mathbb{C}^\dagger$, the total number of independent operators is
given by
\begin{equation}
    (1+8+10) \cdot (6+3) \cdot 3^3 = 4617\,,
\end{equation}
in agreement with line~8 of \cref{lst:smeft:ahem}.


\subsection{GRSMEFT at Mass Dimension 12}\label{sec:dim12gr}

The complete basis of operators up to mass dimension~12 for \grsmeft\ is
provided, in the format described in \cref{sec:norepeat,sec:repeat}, in the
ancillary files together with this paper. As an example, we present in
\cref{lst:dim12gr} the operators of type
\begin{equation}\label[type]{eq:smeft:fiji}
  \begin{aligned}
    C_{\mathrm{L}} L^2 e_\mathbb{C} {d_\mathbb{C}^\dagger}^2 u_\mathbb{C}^\dagger D \,.
  \end{aligned}
\end{equation}

\begin{lstlisting}[caption={Content of the file
    \texttt{1CL\symbol{95}2L\symbol{95}1eC\symbol{95}2dC+\symbol{95}1uC+\symbol{95}1D.yml},
    which encodes the operators of the \cref{eq:smeft:fiji}.},
 label={lst:dim12gr}, style=yaml]
# '1CL_2L_1eC_2dC+_1uC+_1D.yml' generated by AutoEFT 1.0.0
version: 1.0.0
type:
- {CL: 1, L: 2, eC: 1, dC+: 2, uC+: 1, D: 1}
- complex
generations: {CL: 1, L: 3, eC: 3, dC+: 3, uC+: 3}
n_terms: 7
n_operators: 1539
invariants:
  Lorentz:
    O(Lorentz,1): -eps(1_1,2_1)*eps(1_2,3_1)*eps(1_3,3_2)*eps(1_4,4_1)*eps(3_1~,6_1~)*eps(5_1~,7_1~) * CL(1_1,1_2,1_3,1_4)*L(2_1)*(D L)(3_1,3_2,3_1~)*eC(4_1)*dC+(5_1~)*dC+(6_1~)*uC+(7_1~)
    O(Lorentz,2): -eps(1_1,2_1)*eps(1_2,3_1)*eps(1_3,3_2)*eps(1_4,4_1)*eps(3_1~,5_1~)*eps(6_1~,7_1~) * CL(1_1,1_2,1_3,1_4)*L(2_1)*(D L)(3_1,3_2,3_1~)*eC(4_1)*dC+(5_1~)*dC+(6_1~)*uC+(7_1~)
    O(Lorentz,3): +eps(1_1,2_1)*eps(1_2,3_1)*eps(1_3,4_1)*eps(1_4,4_2)*eps(4_1~,6_1~)*eps(5_1~,7_1~) * CL(1_1,1_2,1_3,1_4)*L(2_1)*L(3_1)*(D eC)(4_1,4_2,4_1~)*dC+(5_1~)*dC+(6_1~)*uC+(7_1~)
    O(Lorentz,4): +eps(1_1,2_1)*eps(1_2,3_1)*eps(1_3,4_1)*eps(1_4,4_2)*eps(4_1~,5_1~)*eps(6_1~,7_1~) * CL(1_1,1_2,1_3,1_4)*L(2_1)*L(3_1)*(D eC)(4_1,4_2,4_1~)*dC+(5_1~)*dC+(6_1~)*uC+(7_1~)
    O(Lorentz,5): -eps(1_1,2_1)*eps(1_2,3_1)*eps(1_3,4_1)*eps(1_4,5_1)*eps(5_1~,6_1~)*eps(5_2~,7_1~) * CL(1_1,1_2,1_3,1_4)*L(2_1)*L(3_1)*eC(4_1)*(D dC+)(5_1,5_1~,5_2~)*dC+(6_1~)*uC+(7_1~)
    O(Lorentz,6): +eps(1_1,2_1)*eps(1_2,3_1)*eps(1_3,4_1)*eps(1_4,6_1)*eps(5_1~,6_1~)*eps(6_2~,7_1~) * CL(1_1,1_2,1_3,1_4)*L(2_1)*L(3_1)*eC(4_1)*dC+(5_1~)*(D dC+)(6_1,6_1~,6_2~)*uC+(7_1~)
    O(Lorentz,7): -eps(1_1,2_1)*eps(1_2,3_1)*eps(1_3,4_1)*eps(1_4,7_1)*eps(5_1~,7_1~)*eps(6_1~,7_2~) * CL(1_1,1_2,1_3,1_4)*L(2_1)*L(3_1)*eC(4_1)*dC+(5_1~)*dC+(6_1~)*(D uC+)(7_1,7_1~,7_2~)
  SU3:
    O(SU3,1): +eps(5_1,6_1,7_1) * CL*L*L*eC*dC+(5_1)*dC+(6_1)*uC+(7_1)
  SU2:
    O(SU2,1): +eps(2_1,3_1) * CL*L(2_1)*L(3_1)*eC*dC+*dC+*uC+
permutation_symmetries:
- vector: Lorentz * SU3 * SU2
- symmetry: {CL: [1], L: [2], eC: [1], dC+: [2], uC+: [1]}
  n_terms: 3
  n_operators: 972
  matrix: |-
    [ 0  0  1  1 -1 -1  2]
    [ 0  0  2  2  0  0  0]
    [ 0  0  0  0  2  2  0]
- symmetry: {CL: [1], L: [2], eC: [1], dC+: [1, 1], uC+: [1]}
  n_terms: 2
  n_operators: 324
  matrix: |-
    [ 0  0  1 -1 -1  1  0]
    [ 0  0  2 -2  0  0  0]
- symmetry: {CL: [1], L: [1, 1], eC: [1], dC+: [2], uC+: [1]}
  n_terms: 1
  n_operators: 162
  matrix: |-
    [ 2  2 -1 -1  1  1 -2]
- symmetry: {CL: [1], L: [1, 1], eC: [1], dC+: [1, 1], uC+: [1]}
  n_terms: 1
  n_operators: 81
  matrix: |-
    [ 2 -2 -1  1  1 -1  0]
\end{lstlisting}

As indicated in \cref{tab:sm_fields}, the symbol \verb$CL$ denotes the
left-handed Weyl tensor $C_{\mathrm{L}}$ defined in \cref{eq:sl2c:gr}. From
the examples above, the reader should by now be able to reconstruct all
operators from this listing.


\subsection{Ancillary Files}

The complete set of \smeft\ and \grsmeft\ operators up to mass dimension $12$
with three generations of fermions are published as ancillary files
together with this paper.  In addition, we provide a table with the numbers
of \textit{types}, \textit{terms}, and \textit{operators} contained in a given
family (see \cref{tab:dim12,tab:gr:dim12} as an example) as well as the
associated Hilbert series for each mass dimension.  This information can be
extracted solely from the attached operator files.

The operators are provided in the format described in
\cref{sec:norepeat,sec:repeat}. The files are included in the archives
\verb$smeft_<ndim>.tar.xz$ and
\verb$grsmeft_<ndim>.tar.xz$, where \verb$<ndim>$ is the mass
dimension.\footnote{Note that the file \texttt{smeft\symbol{95}<ndim>.tar.xz}
contains the complete set of \smeft~operators for mass
dimension~\texttt{<ndim>} while \texttt{grsmeft\symbol{95}<ndim>.tar.xz}
contains only operators including the Weyl tensor. The full \grsmeft~basis is
recovered by the union of \texttt{smeft\symbol{95}<ndim>.tar.xz} and
\texttt{grsmeft\symbol{95}<ndim>.tar.xz}.}  They are collected in directories
as
\begin{verbatim}
  operators/<N>/<family>/<type>.yml
\end{verbatim}
where \verb$<N>$, \verb$<family>$, and \verb$<type>$ are the number of fields
$N$, the family and the type of the operator as defined in
\cref{sec:construct}. Concrete examples are given in
\cref{sec:norepeat,sec:repeat,sec:dim12,sec:dim12gr}.  This format makes
it simple to access specific operators in the set.

For \smeft\ up to mass dimension~12, the size of the files amounts to about
442\,MB\@. This number seems to grow roughly with the number of operators, from
which we infer that it will reach 1\,TB at mass dimension~20, and 1~PB at mass
dimension~26.

The Hilbert series provides the number of operators per type and thus constitutes a helpful
check on our results. We have compared our numbers for \smeft\ and \grsmeft\ up to
dimension~12 to the results for the Hilbert
series as obtained by \texttt{ECO}\,\cite{Marinissen:2020jmb} and found full
agreement. We have further performed a number of consistency checks on our
results, for example that no two $\mathcal{K}$ matrices for a particular
operator type contain linearly dependent rows, etc.  An immediate comparison of the
results at lower mass dimension to the existing literature is highly
non-trivial though, due to different representations of the final
basis~\cite{Li:2020gnx,Li:2020xlh,Murphy:2020rsh}.\footnote{This is even true
for the comparison with the earlier results based on the same algorithm,
because further manipulations have been applied to the final results in
\citeres{Li:2020gnx,Li:2020xlh} in order to present them in a more compact
form.} A general conversion tool which would allow such comparisons is
currently under development~\cite{convert}.


\begin{table}[h]
\centering 
\begin{tabular}{crrr}
\toprule
family & types & terms & operators \\
\midrule
${F_{\mathrm{L}}}^{2} {\phi}^{2} F_{\mathrm{R}} {D}^{4} + \mathrm{h.c.}$ & 22 & 422 & 422 \\
${F_{\mathrm{R}}}^{5} {D}^{2} + \mathrm{h.c.}$ & 22 & 78 & 78 \\
${\phi}^{4} F_{\mathrm{R}} {D}^{6} + \mathrm{h.c.}$ & 4 & 96 & 96 \\
\vdots&\vdots&\vdots&\vdots\\
$F_{\mathrm{L}} {\psi}^{4} {F_{\mathrm{R}}}^{2} + \mathrm{h.c.}$ & 128 & 1590 & 61398 \\
${\psi}^{4} {\phi}^{2} F_{\mathrm{R}} {D}^{2} + \mathrm{h.c.}$ & 58 & 4504 & 161772 \\
\midrule
257 & 11942 & 472645 & 75577476 \\
\bottomrule
\end{tabular}
\caption{\smeft~families at mass dimension 12.
The complete table can be found in \texttt{smeft\symbol{95}12.tar.xz} as \texttt{table12.pdf}.}\label{tab:dim12}
\end{table}

\begin{table}[h]
\centering 
\begin{tabular}{crrr}
\toprule
family & types & terms & operators \\
\midrule
$C_{\mathrm{L}} {F_{\mathrm{L}}}^{2} {\phi}^{2} {D}^{4} + \mathrm{h.c.}$ & 8 & 160 & 160 \\
$C_{\mathrm{L}} F_{\mathrm{L}} {F_{\mathrm{R}}}^{3} {D}^{2} + \mathrm{h.c.}$ & 24 & 96 & 96 \\
${C_{\mathrm{L}}}^{2} F_{\mathrm{L}} F_{\mathrm{R}} C_{\mathrm{R}} {D}^{2} + \mathrm{h.c.}$ & 6 & 24 & 24 \\
\vdots&\vdots&\vdots&\vdots\\
${C_{\mathrm{L}}}^{3} {\phi}^{2} {F_{\mathrm{R}}}^{2} + \mathrm{h.c.}$ & 8 & 8 & 8 \\
${F_{\mathrm{L}}}^{2} \phi {\psi^\dagger}^{2} F_{\mathrm{R}} C_{\mathrm{R}} + \mathrm{h.c.}$ & 94 & 150 & 1350 \\
\midrule
460 & 6097 & 70528 & 3936965 \\
\bottomrule
\end{tabular}
\caption{\grsmeft$\ominus$\smeft~families at mass dimension 12.
The complete table can be found in \texttt{grsmeft\symbol{95}12.tar.xz} as \texttt{table12.pdf}.}\label{tab:gr:dim12}
\end{table}


\clearpage


\section{Conclusions}\label{sec:conclusions}

We have evaluated the \smeft\ and \grsmeft\ operator basis with three
generations of fermions up to mass dimension~12. They were obtained by
re-implementing the algorithm of \citeres{Li:2020gnx,Li:2020xlh,Li:2022tec}
into a non-commercial software. Aside from the results obtained in this paper,
we also confirm the completeness of the algorithm up to mass dimension 12.
The operators are provided in the form of searchable and compact ancillary
files.

In future work, we plan to extend the capabilities of \texttt{AutoEFT} in
various respects, for example to keep operators that vanish by
equations-of-motion~\cite{Ren:2022tvi}, as they are needed for the
renormalization of the operators. Furthermore, we plan to implement general
basis transformations in the spirit of
\citere{Falkowski:2015wza,Sannino:2017utc}, which would allow the fully
automated matching of \efts\ to an ultraviolet complete theory, for example by
combining \texttt{AutoEFT} with methods like
\abbrev{UOLEA}\,\cite{Drozd:2015rsp,Summ:2018oko,Kramer:2019fwz,
  Fuentes-Martin:2022jrf} (see also
\citere{Carmona:2021xtq}).


\paragraph{Acknowledgments.}
We would like to thank Svenja Diekmann, Jakob Linder, and \mbox{Maximilian}
Rzehak for constructive discussions.  This research was supported by the
Deutsche Forschungsgemeinschaft (DFG) under grant 400140256 -- GRK 2497: The
physics of the heaviest particles at the LHC, and grant 396021762 -- TRR 257:
P3H -- Particle Physics Phenomenology after the Higgs Discovery.



\bibliographystyle{utphys}

\bibliography{paper}


\end{document}